\RequirePackage{lineno}
\documentclass[aps,reprint,pre,onecolumn,showpacs,floatfix,groupedaddress,longbibliography]{revtex4-1}
\usepackage{graphicx}
\usepackage{bm}
\usepackage{comment} 
\usepackage{amssymb}
\usepackage{color}
\usepackage{amsbsy}      
\usepackage{amsmath}

\begin{document}

\title[A network model of immigration] 
      {A network model of immigration: Enclave formation vs. cultural integration}
      
\author{Yao-Li Chuang}
 \affiliation{Dept. of Mathematics, CSUN, Los Angeles, CA 91330-8313, \\
Dept. of Biomathematics, UCLA, Los Angeles, CA 90095-1766}

\author{Tom Chou}
 \affiliation{Dept. of Biomathematics, UCLA, Los Angeles, CA 90095-1766, \\
 Dept. of Mathematics, UCLA, Los Angeles, CA 90095-1555}

\author{Maria R. D'Orsogna} \email{dorsogna@csun.edu}
 \affiliation{Dept. of Mathematics, CSUN, Los Angeles, CA 91330-8313,\, \\
Dept. of Biomathematics, UCLA, Los Angeles, CA 90095-1766}

\date{\today}




\begin{abstract}
\noindent
Successfully integrating newcomers into native communities has become 
a key issue for policy makers, as the growing number of 
migrants has brought cultural diversity, new skills, and
at times, societal tensions to receiving countries.  We develop an
agent-based network model to study interacting ``hosts" and ``guests"
and identify the conditions under which cooperative/integrated or
uncooperative/segregated societies arise.  Players are assumed to seek
socioeconomic prosperity through game theoretic rules that shift
network links, and cultural acceptance through opinion dynamics.  We
find that the main predictor of integration under given initial
conditions is the timescale associated with cultural adjustment
relative to social link remodeling, for both guests and hosts.  Fast
cultural adjustment results in cooperation and the establishment of
host-guest connections that are sustained over long times.
Conversely, fast social link remodeling leads to the irreversible
formation of isolated enclaves, as migrants and natives optimize their
socioeconomic gains through in-group connections.  We discuss how
migrant population sizes and increasing socioeconomic rewards for
host-guest interactions, through governmental incentives or by
admitting migrants with highly desirable skills, may affect the
overall immigrant experience.
\end{abstract}

\maketitle

\section{Introduction}
Migrating human populations have always played a significant role in
history \cite{BOY13,CAS03,CRA12}. 
For centuries individuals driven by adventurous spirits, 
or seeking better socio-economic opportunities, have voluntarily abandoned 
their original environments.
Large groups of people have also been
involuntarily forced from their homelands by hostile events such as
famine, drought, religious persecution, political turmoil, human
rights violations, and wars.  According to the United Nations High
Commissioner for Refugees (UNHCR), the number of forcibly displaced
persons worldwide has been steadily climbing since 2011, reaching an
unprecedented level of 68.5 million persons by the end of 2017. 
Among these 28.5 million are asylum seekers or refugees \cite{UNH18}.  
Economic disparity enhances the pull of populations
towards more developed regions; increased mobility reduces the cost in
crossing national borders and geographic barriers; advanced
communication technologies facilitate long distance social
connections. All of these factors contribute to the massive scale of
human migration observed in recent years \cite{CHA79}.

While large-scale emigration causes brain drain and loss of labor
force in ``source" countries, regions receiving immigrants also face
challenges in accommodating new arrivals who may follow different social,
cultural, and religious norms. 
Mistrust between natives and migrants may arise and exacerbate over time
due to inadequate infrastructure and assistance programs.
A well-documented phenomenon among
immigrants is that of acculturative stress \cite{BER92,BER87}, whereby
contact with another culture may lead to psychological and
somatic health issues.  Overall various studies of the immigrant 
experience describe outcomes ranging from very positive to very 
negative \cite{BER92,BER05,IRE04,SEM81}.  Immigrants joining a multicultural
society generally suffer from the least acculturative stress and are
the best adapted, whereas those settling in less culturally tolerant
communities face more challenges \cite{BER05,IRE04}. A common
observation is that those who do not adapt well, either by
circumstances or lack of motivation, often become socially
marginalized.  Self-segregation may lead to the creation of insular
communities that offer advantages to immigrants, 
but that also prevent them from fully integrating \cite{BER92,BER05,SEM81}.  
These enclaves often deepen divisions between host and immigrant groups.
The attitude of the majority host population is an important predictor
of how successful the adaptation process of an immigrant group will
be.  Hostile host communities tend to hinder adaptation, with averse
majorities playing a key role in the emergence of segregated minority
communities \cite{KOO10,PRI14}.

The complex relationship between natives and migrants evolves over
time and depends on many economic, historic, and political factors.
By framing the main ingredients of this relationship in simple,
quantifiable ways, mathematical models may help one to understand the
implications of various mechanisms and of their synergy, and may help
design intervention strategies.  Agent-based mathematical models have
been recently employed to study coexistence and cooperation among
culturally heterogeneous populations through game theory
\cite{CHI13,COH01,FEH11,HAL00,HAM06b,KLO99,NEM07,RIO97,RIO01,WAN18}, opinion
dynamics \cite{CHU17,DEF00,DEG74,FEL13,FEL14,FRI99,GOL10,KRA00,WEI02},
population dynamics \cite{AXE97,FOS06,HAW18}, and network theory
\cite{COH01,FEH11,HAM06b,HEN11}.  In this paper we introduce an
agent-based social-network model that assumes immigrant groups have
two primary objectives: to improve their socioeconomic status and to
gain acceptance within their social circles.  The former scenario is
usually modeled by implementing game theory rules, whereby a utility
function associated with socioeconomic status is to be maximized
\cite{CHI13,COH01,HAL00,HAM06b,NEM07,RIO01}. The latter is typically
described using opinion dynamics, whereby individuals adjust their
opinions or cultural traits through social interactions
\cite{CHU17,DEF00,DEG74,FEL13,FEL14,FRI99,GAL05,KRA00,WEI02}.  Simplistic
game theory models rarely yield cooperative patterns, as defectors
tend to prevail if each agent is allowed to only make rational
decisions for his or her own self-interest \cite{NEM07}.  Cooperative
behavior may emerge through biased decision making whereby individuals
collaborate solely with those that share their same opinion.  This
mechanism leads to social segregation, as tight collaborations develop
only within culturally homogeneous enclaves
\cite{CHI13,COH01,FEH11,HAL00,HAM06b,NOW90,RIO97,RIO01}.  
Models of opinion dynamics on the other hand often assume individuals seek
like-minded peers, and willingly adjust to prevailing opinions
\cite{AXE97,GAL05}. Minority opinions arise and persist only through ad-hoc
restrictions, such as including zealots,
or by imposing thresholds so that consensus
is reached only if two opinions are sufficiently close \cite{GAL16a,NOW90}.

As a rule of thumb, game theoretic models
result in uncooperative behavior;
opinion dynamics leads to uniform consensus.  The
immigrant narrative, however, is much more nuanced with behaviors
ranging from uncooperative segregation to cooperative integration, 
suggesting modeling should include both mechanisms.
We thus introduce a network populated with interacting ``guest" and ``host"
nodes that seek to improve their socioeconomic status
while culturally adjusting to each other.  Socioeconomic gains are
modeled via a utility function that evolves through game theoretic
rules, while the attitudes (or ``opinions") that players
harbor towards others evolve through opinion dynamics. 
These two mechanisms are interdependent, so
that attitudes towards different cultures shape utility gains, and vice versa.

We show that the main predictor of integration or segregation is given
by the relationship between two timescales: that of cultural
adjustment, whereby guests and hosts adapt more tolerant attitudes of
each other, and that of social link remodeling, whereby players change
their network connections to increase their socioeconomic rewards. In
the case of slow cultural adjustment, immigrant and host communities
tend to segregate as accumulation of socioeconomic wealth occurs more
efficiently through insular in-group connections.  Conversely, if
adjustment is sufficiently fast, cross-cultural bridges may be
established and sustained, allowing different cultural groups to reach
``consensus" and maintain active cooperation.  Another key role will
be played by the fraction of immigrants joining the total population
as the immigrant-to-host ratio changes the cultural adjustment timescales. 
As we outline below, a high immigrant ratio increases
the likelihood of in-group connections and reduces communication
between immigrant and host populations.

In Section\,\ref{SEC:MODEL}, we introduce our network model, the
mechanisms that govern the evolution of social
connections, and
the utility function for immigrant-host interactions.  In
Section\,\ref{SEC:RESULTS} we examine the parameter dependence of our
model and show how processes unfolding over different timescales
lead to different
outcomes of immigration integration. Finally, we
conclude in Section \ref{SEC:DISCUSSION} with a discussion on
sociological and policy implications.

\section{The model\label{SEC:MODEL}}

\noindent
Our basic model consists of a network
whose nodes symbolize immigrant or
native agents connected by edges that represent social links.  Each
node is also associated with an attitude and a utility function that
depend on its connections and that determine an agent's socioeconomic
status. Over time, nodes change their connections and attitudes as
they seek to increase their utility; as a result the network evolves
towards integration or segregation between immigrant and host
communities.

\subsection{Network}

\noindent
Within our network model a node represents a social unit, such as
an individual or a collection of individuals,
and is labeled as a ``guest'' or a
``host'', depending on whether it belongs to the immigrant or native
group.  Each node, indexed by $i$, is characterized by an ``attitude''
variable $x^t_i$ at time $t$, which varies between
$-1 \le x^t_i \le 0$ for guest nodes and in
$0 \le  x^t_i \le 1$
for host nodes. Hence
the sign of $x_i^t$ is used to distinguish the group identity of the
node. The magnitude $\vert x^t_i \vert$ indicates the degree of
hostility that node $i$ harbors towards those belonging to the other
group. Thus,
$x^t_i \to 0^{\pm}$
characterizes most receptive guests or most
hospitable hosts, while $x^t_i = \pm 1$ represents the highest level
of xenophobia. 
Moreover, we define $\Omega_i^t$ as the
``social circle''
  of node $i$ at time $t$, which is a set containing
all nodes directly connected to node $i$ at time $t$. 
We assume
that there are a fixed number of $N_{\rm h}$ host and $N_{\rm g}$
guest nodes, with varying
attitudes. 
All nodes $N = N_{\rm h}
+ N_{\rm g}$ seek to maximize their utility function as defined below.

%
\begin{figure}[t]
  \begin{center}
      \includegraphics[width=2.5in]{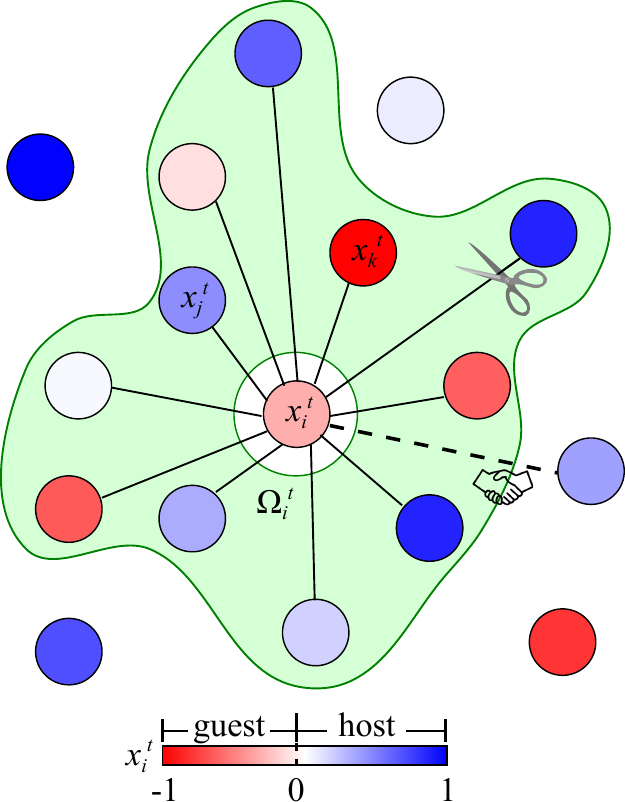}
\end{center}
  \caption{Model diagram. Each node $i$ is characterized by a variable
    attitude $-1 \le x_i^t \le 1$ at time $t$. Negative values,
    depicted in red, indicate guest nodes; positive values represent
    hosts, colored in blue.  The magnitude $\vert x_i^t \vert$
    represents the degree of hostility of node $i$ towards members of
    the other group. Each node is shaded accordingly.  All 
    nodes $j,k$ linked to the central node $i$ 
    represent the green-shaded social circle
    $\Omega_i^t$ of node $i$ at time $t$. The utility $U_i^t$ 
    of node $i$ depends on its attitude relative to that of
    its $m^t_i $ connections in $\Omega_i^t$ and on $m^t_i$. 
    Nodes maximize their utility by adjusting
    their attitudes $x_i^t$ and by establishing or severing
    connections, reshaping the network over time.
    \label{FIG:MODEL}}
\end{figure}

\subsection{Utility Function}

\noindent
The dynamics of our network is driven by the utility function $U_i^t$
assigned to each node $i$. Each player seeks to maximize $U_i^t$ by
shifting its attitude $x_i^t$, and by forging and severing connections
with other nodes. We model the utility $U_i^t$ of node $i$ at time $t$
via two components: a reward function $u^t_{ij}$ for interacting with
node $j$, and a cost function $c (m_i^t)$ for maintaining
$m_i^t$ connections so that

\begin{eqnarray}
  U_i^t & = & \sum_{j \in \Omega_i} u^t_{ij} - c (m_i^t)
  \label{EQ:UTILITY} \\
        & = & \sum_{j \in \Omega^t_i} A_{ij}
    \exp \left( - \frac{\left( x^t_i - x^t_j \right)^2}{2 \sigma} \right)
    - \exp \left( \frac{m_i^t}{\alpha} \right).
    \nonumber
\end{eqnarray}

\noindent
The pairwise reward function $u^t_{ij}$ depends on the attitude
difference $\vert x^t_i - x^t_j \vert$ between
connected nodes $i$ and $j$;
the smaller the attitude difference, the higher the reward. For a pair
of nodes from the same group, \textit{i.e.} if both $i$ and $j$ are
hosts or immigrants, maximizing $u^t_{ij}$ implies $x_i^t = x_j^t$
leading to consensus within the group. If $i$ and $j$ are nodes from
different groups, $u^t_{ij}$ is maximized by both sides adopting 
more cooperative attitudes such that $x_i^t \to 0^-$ and $x_j^t \to 0^+$. 
Hence, the value of $x_i^t$ that will maximize $U_i^t$ will depend on the
composition of $\Omega_i^t$ and the attitudes $x^t_j$ of its members.
The parameter $\sigma$ controls the sensitivity of the reward, while
the amplitude $A_{ij}$ specifies the maximum reward attainable when
$x_i^t = x_j^t$. In principle, $A_{ij}$ may depend on the specific
socioeconomic attributes of the interacting $i, j$ pair. 
For simplicity we let $A_{ij}$ be one of two discrete levels; $A_{ij} =
A_{\rm in}$ for in-group interactions, where nodes $i$ and $j$ belong
to the same group, both hosts or both migrants, and $A_{ij} = A_{\rm
  out}$ for out-group interactions between nodes $i$ and $j$ of
different groups. The cost
function $c$ in Eq.\,\ref{EQ:UTILITY} is a function of $m_i^t =
\left\vert \Omega^t_i \right\vert$, the number of connections
sustained by node $i$ at time $t$, which by definition is also the
cardinality of the social-circle set $\Omega^t_i$.  We assume that the
cost to maintain connections increases exponentially with $m_i^t$
through a scaling coefficient $\alpha$. A smaller $\alpha$ value
results in a steeper increase of cost, leading to fewer average
connections per node.  
Note that such a cost function penalizes nodes with too
many connections, suppressing the likelihood of ``hub'' nodes of high
connectivity, a hallmark of small world networks that characterizes
many real world social networks. In more realistic settings, the cost
of maintaining social connections depends on more nuanced characteristics
of each individual (wealth, fame, age, community status), allowing
some to sustain higher degrees of connectivity than others. For
simplicity our model does not include these considerations.

\subsection{Mechanisms of Model Evolution}
\label{mech}

\noindent
At each time step, each node $i$ seeks to increase its utility $U_i^t$
by adding or cutting connections and adjusting its attitude $x_i^t$.
We model this process as a series of stochastic events through the
following steps:

\begin{enumerate}
\item At time $t$, randomly pick the ``active" node $i$
    to make
    a decision.
    \label{STEP1}
\item Randomly pick another node $j \ne i$.
  \begin{itemize}
  \item If $i$ and $j$ are connected, i.e., $j \in
    \Omega_i^t$, check whether breaking the $i$--$j$ connection
    increases $U_i^t$ for node $i$.  If it does, break the $i$--$j$ connection.
  \item If $j \notin \Omega_i^t$, check whether adding an $i$--$j$
    connection increases $U_i^t$ for node $i$. If it does, add the $i$--$j$
    connection.
  \end{itemize}
\item Randomly pick a connected node
  $\ell \in \Omega_i^t$ via a reward-weighted
  probability
  \begin{equation}
    p_{\ell} = \frac{u_{i \ell}^t}{\sum_{\ell \in \Omega_i^t} u_{i \ell}^t}. \label{EQ:PROBABILITY}
  \end{equation}
\item Determine $x_i^{t+1}$ using $x_i^{t}$, $x_{\ell}^t$ 
  \begin{equation}
    x_i^{t+1} =   \left\{
    \begin{array}{ll}
      \min \left( \displaystyle 0, x_i^{t} + \frac{x_{\ell}^t - x_i^t}{\kappa} \right) 
      & \textrm{ for } x_i^{t} < 0 \textrm{ (guest)}, \\
      \max \left( \displaystyle 0, x_i^{t} + \frac{x_{\ell}^t - x_i^t}{\kappa} \right) 
      & \textrm{ for } x_i^{t} > 0 \textrm{ (host)}, \\
      \end{array} \right.
          \label{EQ:CONFORMATION}
  \end{equation}
   where $\kappa$ is the timescale associated with attitude adjustment. 
   Large values of $\kappa$ indicate longer adaptation times.  We select different
  nodes $j \neq \ell$ for remodeling network connections and adjusting
  attitudes to avoid the emergence of any systematic biases. \label{STEPF}
\item Advance time $t \to t + (1/N)$ and repeat steps \ref{STEP1}--\ref{STEPF}.
\end{enumerate}

\noindent
In the above steps, all unweighted random selections are made through
a uniform probability. As presented, our algorithm alternates between
remodeling network connections and making attitude adjustments.  Note
that when steps \ref{STEP1}--\ref{STEPF} are repeated on all $N$ nodes, $t$ 
advances to $t + 1$, and that, on average, each node makes
decisions once within this unitary time step.  Thus, the timescale for
network remodeling is one.  The timescale for attitude adjustment,
instead, is given by $\kappa$ scaled by the probability for node $i$ to
be paired with node $\ell$ carrying a different attitude. We can
approximate this probability as the fraction of out-group connections,
$N_{\rm g}/N$ for hosts and $N_{\rm h}/N$ for guests, so that the 
guest adjustment timescale $\tau_{\rm h}$
can be estimated by $\tau_{\rm h} \sim \kappa N/N_{\rm g}$, 
and the host adjustment timescale $\tau_{\rm g}$ by 
 $\tau_{\rm g} \sim \kappa N/N_{\rm h}$.

An important observation is that $U_i^t$ can reach its maximum
$U_i^{\rm max}$ if $|x^t_i - x^t_j| \to 0$ within connected components
of the network. This can be achieved in two different ways: i) through
actual consensus where all nodes carry a neutral attitude $x^t_i \to
0$ so that in-group and out-group connectivities are equally likely,
or ii) through a segregated network with homogeneous clusters made of
all guests or all hosts, where non-zero but uniform attitudes are
maintained in each cluster, so that $|x_i - x_j^t| \to 0$ does not
necessarily imply $x^t_i \to 0$. Although these two different network
configurations lead to the same maximal utility, only the first one
will be considered a true hallmark of harmonious integration, since
attitudes are the most open on both sides, and there is minimal
differentiation between intra-group or out-group connectivity.  The
second case instead represents the creation of parallel societies,
with each group self-segregating into its own homogeneous enclave,
maintaining little contact with ``the other".

\subsection{Initial Conditions}
\label{SEC:IC}

\begin{table}[t]
\begin{tabular}{|c|c|c|}
\hline
\hline
Symbol & Description & default values\\
\hline
\hline
 $x_i$ & \hspace{4cm} attitude \hspace{4cm} & -1 to 1  \\ 
\hline
 $A_{\rm in}$ & maximal utility through in-group connection & $10$   \\ 
\hline
 $A_{\rm out}$ & maximal utility through out-group connection &
 $1$ to $100$  \\ 
\hline
 $\sigma$ &  sensitivity to attitude difference & $1$  \\ 
\hline
 $\kappa$ &  attitude adjustment timescale & $100$ to $1000$ \\
\hline
 $\alpha$ & cost of adding connections & $3$ \\
\hline
 $N$ & total population & $2000$ \\
\hline
$N_{\rm g}$ & guest population & $20$ to $200$ \\
\hline
$N_{\rm h}$ & host population & $N - N_{\rm g}$ \\
\hline
\hline
\end{tabular}
\begin{center}
\caption{List of variables and parameters of the model.
\label{TABLE:MODEL}}
\end{center}
\end{table}

\noindent
All model parameters and typical values are listed in Table
\,\ref{TABLE:MODEL}. Unless otherwise specified, our network
simulations are  performed using the initial conditions described here.
We mostly simulate $N = 2000$ nodes, within which $N_{\rm h} = 1800$
are hosts and $N_{\rm g} = 200$ are guests. In Section
\ref{SEC:TEMPORAL} we also simulate the setting of $N_{\rm g} = 20$
and $N_{\rm h} = 1980$ to examine the effect of extremely small
fractions of guests.  The initial attitudes are set at $x_i^0 = 1$ for
all host nodes, and $x_i^0 = -1$ for all guest nodes, assuming that
before the two groups make any contact they have minimal knowledge on
how to coexist.  For initial connections, we mostly use the following
two extreme and opposite scenarios. One is that host and guest nodes
are randomly connected with uniform probability, yielding on average
ten connections per node at $t=0$. The other is that hosts are
connected to each other and that no guests are present. Host
connectivity is determined by allowing the system to equilibrate in
the absence of guests, representing the natural state of the community
before the arrival of immigrants. Guests are introduced at $t=0$ as
nodes without any links to either hosts or fellow guests.  Note that
because of the definition of the utility function in
Eq.\,\ref{EQ:UTILITY}, and because we allow the host community to
equilibrate prior to inserting guests, we expect each host to be
connected to an average number of $\alpha \ln \left( \alpha A_{\rm in}
\right)$ other hosts at $t=0$.  The first initial condition scenario
represents a perfectly executed welcoming program for immigrants, 
providing with them sufficient social ties to connect to the native 
community.  In the second initial condition scenario, such a welcoming 
program does not exist at all, and guests arrive in a completely foreign 
environment.

\section{Results\label{SEC:RESULTS}}

\begin{figure}[t]
  \begin{center}
      \includegraphics[width=5.0in]{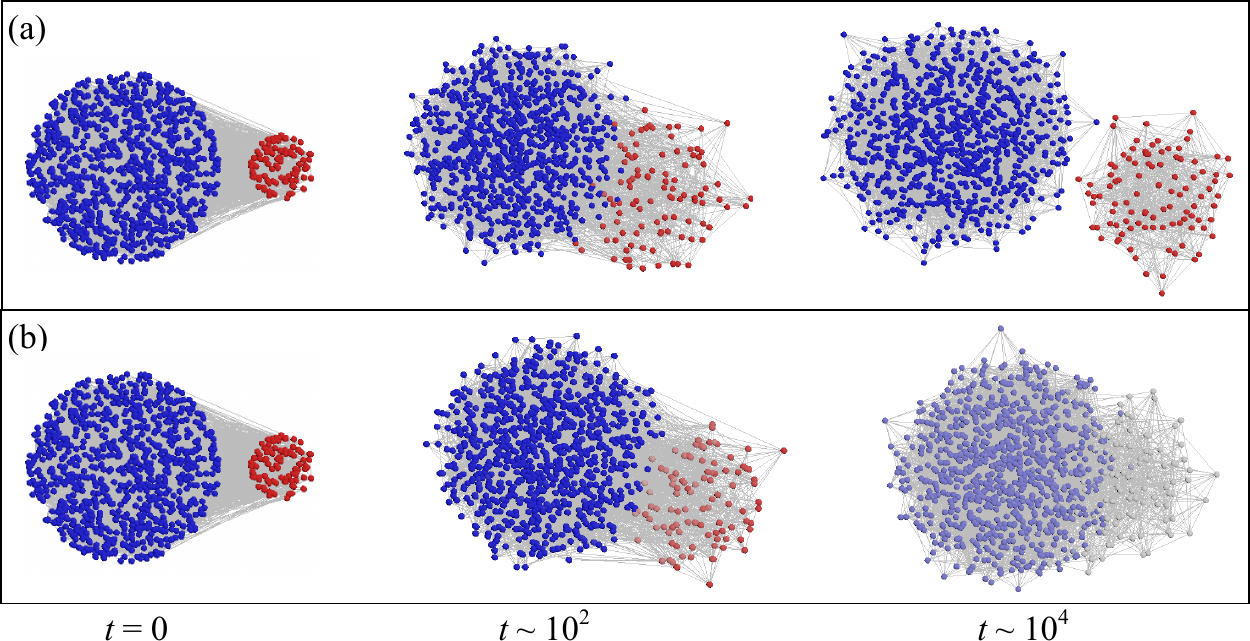}
\end{center}
  \caption{Simulated network dynamics leading to (a) complete segregation, and
    (b) integration between guest (red) and host (blue)
    populations. Shading of node colors represents the degree of hostility
    $|x_i^t|$ of node $i$ towards those of its opposite group,
    according to the color scheme shown in
    Fig.\,\ref{FIG:MODEL}. Initial conditions are randomly connected
    guest and host nodes with attitudes $x_{i,{\rm guest}}^{0} = -1$
    and $x_{i,{\rm host}}^{0} = 1$.  Other parameters are $N_{\rm
      h}=900, N_{\rm g}=100$, $\alpha = 3$, $A_{\rm in} = A_{\rm out}
    = 10$, $\sigma = 1$. The two panels differ only for $\kappa$, the
    attitude adjustment timescale, with $\kappa = 1000$ in panel (a)
    and $\kappa = 100$ in panel (b).  (a) For slowly changing attitudes
    ($\kappa = 1000$), hostile attitudes persist over time, eventually
    leading to segregated clusters. (b) For fast changing attitudes
    ($\kappa = 100$), guests initially become more cooperative, as
    shown by the lighter red colors. Over time, a more connected
    host--guest cluster arises with hosts eventually adopting more
    cooperative attitudes as well.
    \label{FIG:SNAPSHOT}}
\end{figure}

\noindent
Figure\,\ref{FIG:SNAPSHOT} shows two representative outcomes of our
network model at steady state.  In Fig.\,\ref{FIG:SNAPSHOT}a guests
(red circles) and hosts (blue circles) segregate and maintain highly
hostile attitudes, as illustrated by the dark red and blue shades of
the right hand panel.  Cross-group utilities at the beginning of simulations 
yield low rewards which do not increase over time, leading to the severing of
all ties between hosts and guests at $t \to \infty$. In
Fig.\,\ref{FIG:SNAPSHOT}b guests adopt more cooperative attitudes as
represented by the lighter red colors.  Such attitudes increase
cross-group rewards so that guests and hosts stay mixed.  Hosts will
also become more cooperative, although at slower timescales than
guests.
%
%

The two configurations shown in Fig.\,\ref{FIG:SNAPSHOT}
represent two ways through which $U_i^t$ in
  Eq.\,\ref{EQ:UTILITY} is maximized.
  The configuration in Fig.\,\ref{FIG:SNAPSHOT}a arises
  by cutting all cross-group links to form enclaves, within which
  guests and hosts adopt uniform but different attitudes
  $x_{i,{\rm guest}} \neq x_{i,{\rm host}} \neq 0$.
  The configuration in Fig.\,\ref{FIG:SNAPSHOT}b emerges through
  cooperative attitudes $x_{i,{\rm guest}} = x_{i,{\rm host}} = 0$ for
  all players.
Both lead to $\vert
x_i^t - x_j^t \vert \to 0$ as $t \to \infty$.  To which 
of these two basins of attraction society converges, 
will depend on parameter choices and initial conditions as discussed below.

\subsection{Maximizing utilities via network remodeling 
and attitude adjustment}
\label{SEC:TEMPORAL}

\noindent
For a more quantitative perspective, we now examine how the utility
function $U_i^t$ and the attitude profiles $x_i^t$ vary over time in
some sample simulations.  We set the model parameters to $\alpha = 3$,
$A_{\rm in} = A_{\rm out} = A = 10$, and $\sigma = 1$, and let
$\kappa$ vary between $100$ and $1000$ with $N=2000$ and $N_{\rm g}
=200$ or $N_{\rm g} =20$.  The assumption $A_{\rm in} = A_{\rm out} =
A$ leads to a maximum in the utility $U_i^t = U_i^{\rm max} = \alpha A
[\ln (\alpha A) - 1]$ which is reached if all connected nodes conform
their attitude so that $\vert x_i^t - x_j^t \vert \to 0$ for any
linked $i,j$ pair and when each node $i$ has $m^t_i = m_{\rm opt} =
\alpha \ln (A_{\rm in} \alpha)$ links.  For our chosen parameters,
$m_{\rm opt} = 10$ connections and $U_i^{\rm max} = 72$.

Since we are interested in how immigrants adapt to their host
environment, we will mainly focus on quantities associated with guest
nodes.  Although host node properties will also dynamically evolve, 
relative changes to their attitudes $x_i^{t}$ and connections will be
much slower than that of guests due to their overwhelming majority.
Initial conditions are chosen so that guests and hosts are randomly
connected to each other as described in Section \ref{SEC:IC}.  For
$N_{\rm g} = 200$ the relatively large number of guests allows for
segregated clusters to emerge and persist with $m_{\rm opt} = 10$
in-group connections.  For $N_{\rm g} = 20$ 
the low number of guests either leads to smaller in-group
guest clusters with less-than-optimal number of connections 
($m_{\rm opt} < 10$), or forces host-guest mixture to reach 
$m_{\rm opt} = 10$.
We will first examine
network remodeling and attitude adjustment independently of each other,
and later the interplay between the two mechanisms.

\begin{figure}[t]
  \begin{center}
      \includegraphics[width=5.0in]{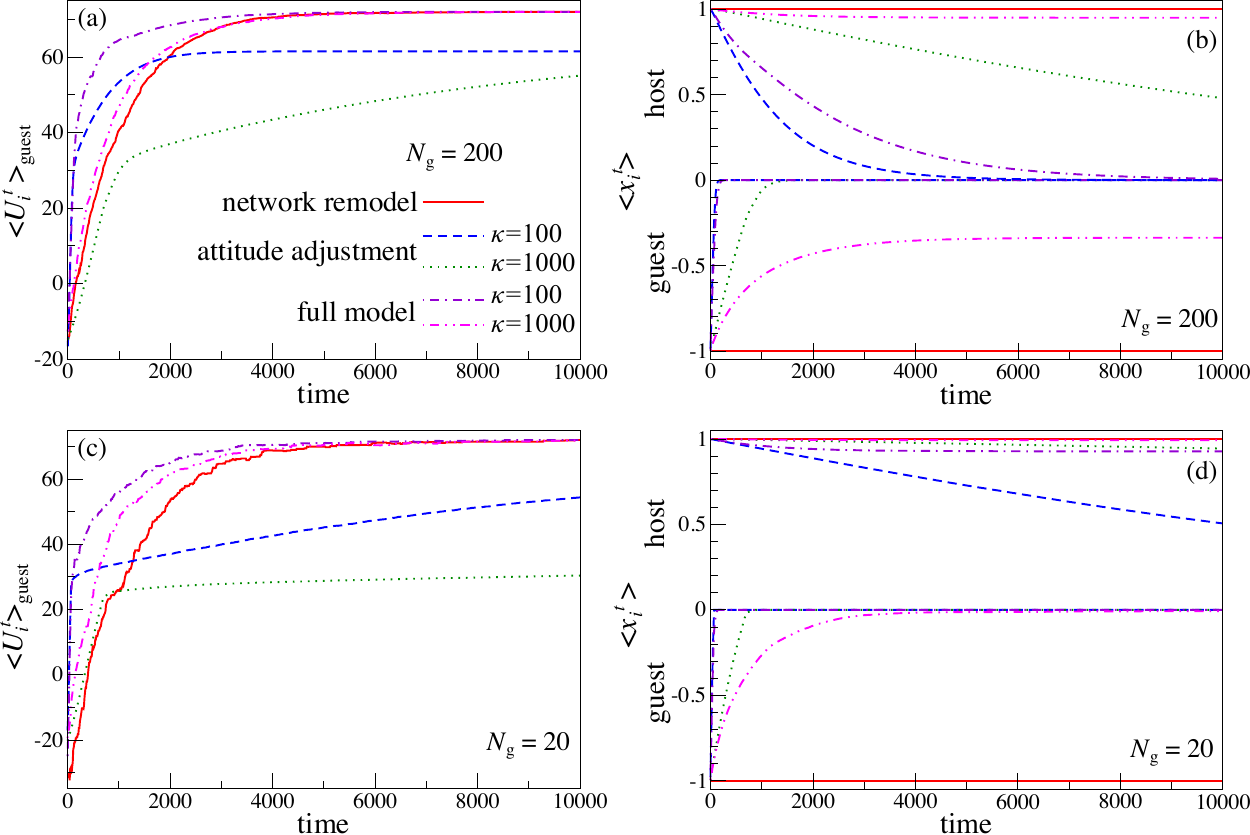}
\end{center}
  \caption{Dynamics of the average utility per node $\langle U_i^t
    \rangle_{\rm guest}$ in panels (a) and (c), and of the average
    attitudes $\langle x^t_{i} \rangle_{\rm guest}, \langle x^t_{i}
    \rangle_{\rm host}$ in panels (b) and (d) for $N_{\rm g} = 200$
    (a,b) and $N_{\rm g} = 20$ (c,d) guests in a total population of
    $N=2000$ nodes.  Parameters are $\alpha = 3$, $A_{\rm in} =A_{\rm
      out} = 10$, and $\sigma = 1$, and $\kappa = 100$ (faster) and
    $\kappa =1000$ (slower) attitude adjustment.  Initial attitudes
    are $x_{i,{\rm{host}}}^0 = 1$ and $x_{i,{\rm {guest}}}^0 = -1$,
    with random connections between nodes so that on average each node
    is connected to $m_i^0=10$ others at $t = 0$, representing full
    insertion of guests into the community.  Network remodeling
    (solid-red curve) and attitude adjustment (blue-dashed and
    green-dotted curves) are considered separately; their interplay is
    illustrated in full model simulations (purple-dot-dashed and
    magenta-double-dotted-dashed).  Utility is increased in all cases,
    but attitude adjustment is more efficient at the onset due to the
    initially set cross-group connections.  Network remodeling allows
    for higher utilities at longer times.  For the full model, fast
    adjustment ($\kappa = 100$) leads to well integrated societies for
    $N_{\rm g} = 200$ as $t \to \infty$, given that 
    $\langle x^t_{i} \rangle_{\rm host} \to 0^{+}$ 
    and $\langle x^t_{i} \rangle_{\rm guest} \to 0^{-}$; for
    $N_{\rm g} = 20$ hosts and guests segregate, with guests adopting
    collaborative attitudes, $\langle x^t_{i} \rangle_{\rm host} \to
    0.93$ and $\langle x^t_{i} \rangle_{\rm guest} \to 0^{-}$.  Under
    slow adjustment ($\kappa = 1000$) hosts and guests will remain
    hostile and segregated with $\langle x^t_{i} \rangle_{\rm host}
    \to 0.95$, $\langle x^t_{i} \rangle_{\rm guest} \to -0.34$ for
    $N_{\rm g} = 200$ and $\langle x^t_{i} \rangle_{\rm host} \to
    0.99, \langle x^t_{i} \rangle_{\rm guest} \to 0^-$ for $N_{\rm g}
    = 20$.
  \label{FIG:TEMP_EVOL}}
\end{figure}

Figure \ref{FIG:TEMP_EVOL} shows the temporal evolution of the average
utility $\langle U^t_{i} \rangle_{\rm guest}$ per guest node and the
average attitude of guests and hosts $\langle x^t_{i} \rangle_{\rm
  guest}$ $\langle x^t_{i} \rangle_{\rm host}$ for sample simulations
of $N_{\rm g} = 200$ (Figs.\,\ref{FIG:TEMP_EVOL}a and
\ref{FIG:TEMP_EVOL}b) and $N_{\rm g} = 20$
(Figs.\,\ref{FIG:TEMP_EVOL}c and \ref{FIG:TEMP_EVOL}d) guest nodes
with $N =2000$ total nodes.  In the red-solid curves we only allow for
network remodeling, and deactivate attitude adjustment.  Vice-versa,
in the blue-dashed ($\kappa = 100$) and green-dotted curves ($\kappa =
1000$) we only allow for attitude adjustment and deactivate network
remodeling.  Finally, the purple dotted-dashed curve ($\kappa = 100$)
and the magenta-double-dotted-dashed curve ($\kappa = 1000$) are
results from the full model, where both network remodeling and
attitude adjustment are implemented.

As can be seen in Fig.\,\ref{FIG:TEMP_EVOL}a for $N_{\rm g} = 200$,
$\langle U^t_{i} \rangle_{\rm guest}$ increases over time towards
$U_i^{\rm max} = 72$ for all five chosen cases. When only network
remodeling is allowed (red-solid curve), $\langle U^t_{i} \rangle_{\rm
  guest}$ increases quickly at the onset of the dynamics as nodes
efficiently exchange low-utility, out-group connections for
high-utility, in-group ones.  As the number of exchanges nears
completion, $\langle U^t_{i} \rangle_{\rm guest}$ increases at a
slower rate, until it converges to the steady state at $U^{\rm
  max}_{i} = 72$ with optimal, high-utility connections that are
mostly in-group.  Guests have established their own self segregated
communities and thrive within it.  When only attitude adjustment is
activated (blue-dashed $\kappa = 100$ and green-dotted $\kappa = 1000$
curves), nodes can only change their attitude and not their
connections, hence they tend to evolve towards conformity ($\vert
x_i^t - x_j^t \vert \to 0$ for all nodes $i,j$).  Note that if $i,j$
are a guest-host pair respectively, conformity will only arise from
$x_{i,{\rm{host}}} \to 0^{-}, x_{j,{\rm{host}}} \to 0^{+}$.  Since
$\langle U^t_{i} \rangle_{\rm guest}$ depends solely on attitude
adjustment, its dynamics will vary on the same timescale as $x_i^t$,
given by $\tau_{g} = N/N_{\rm h} \kappa$.  In the case of fast
attitude adjustment (blue-dashed curve for $\kappa = 100$), the early
rise of $\langle U^t_{i} \rangle_{\rm guest}$ can be more pronounced
than in the case of network remodeling (red-solid curve), as can be
seen for short times ($t \lesssim 2000$) in
Fig.\,\ref{FIG:TEMP_EVOL}a.  However, the utility at steady state
$\langle U^{\rm ss}_{i} \rangle_{\rm guest}$ under attitude adjustment
is lower than under network remodeling, regardless of $\kappa$.  This
is because when only attitude adjustment is allowed, network
connections cannot be rearranged, resulting in a less-than-optimal
connectivity that changes to $x_i^t$ can only partially alleviate.
Having network adjustment as the sole mechanism at play allows for
more flexibility, since, although $x_i^t$ cannot change, a given node
can actively search for others with similar attitude and even increase
its number of connections. We verified that when only one of the two
mechanisms is allowed, attitude adjustment consistently leads to less
optimal outcomes compared to network remodeling for a number of
parameter choices and initial conditions.

%
%

These trends are confirmed and better elucidated by inspecting the
average attitudes of guests $-1 \le \langle x^t_{i} \rangle_{\rm
  guest} \le 0$ and hosts $0 \le \langle x^t_{i} \rangle_{\rm host}
\le 1$ as a function of time in Fig.\,\ref{FIG:TEMP_EVOL}b.  We use
the same parameter sets and initial conditions as in
Fig.\,\ref{FIG:TEMP_EVOL}a and the same color-coding scheme.  The
red-solid curves correspond to the case where we only allow for
network readjustment and attitudes stay unmodified, so $\langle x_i^t
\rangle_{\rm guest} = -1$ and $\langle x_i^t \rangle_{\rm host} = 1$
for all times. The blue-dashed and green-dotted curves, where only
attitude adjustment is allowed show that as $t$ increases,
$\langle x_i^t \rangle_{\rm guest} \to 0^-$ at a faster rate, 
and that $\langle x_i^t \rangle_{\rm host} \to 0^+$ at a much slower one.  
This is easily understood. Since nodes are not allowed to rewire their 
connections, they can only adapt their attitudes as discussed above, 
and provided the network is connected and no isolated clusters exist, all nodes
will eventually conform to $x_i^t \to 0$.  However, being a numerical
minority in the network, guests, for which $x_{i,{\rm {guest}}}^t \le
0$, will share a large number of connections with hosts, for which
$x_{\ell, {\rm {host}}}^t \ge 0$. Under this condition, the adaptation
rules presented in Sec.\,\ref{mech} drive guests towards conformity
more than hosts, so that $\langle x_i^t \rangle_{\rm guest} \to 0^{-}$
faster than $\langle x_i^t \rangle_{\rm host} \to 0^{+}$.  Hence, the
early increases in $\langle U_i^t \rangle_{\rm guest}$ when only
attitude adjustment is allowed and observed in
Fig.\,\ref{FIG:TEMP_EVOL}a (blue-dashed $\kappa = 100$, and
green-dotted $\kappa=1000$ curves) can be attributed to fast
adaptation of guests with time scale $\tau_{\rm g} = \kappa N / N_{\rm
  h}$, and the later increases to slow adaptation of hosts with time
scale $\tau_{\rm h} = \kappa N / N_{\rm g} \gg \tau_{\rm g}$.

The dynamics of the full model (purple-dotted-dashed $\kappa = 100$,
and magenta-double-dotted-dashed $\kappa = 1000$ curves) depend on the
interplay between the two mechanisms at play, attitude adjustment and
network remodeling, and the respective timescales in gaining utility.
From Fig.\,\ref{FIG:TEMP_EVOL}a, $\langle U^t_{i} \rangle_{\rm guest}$
for the full model with fast attitude adjustment
(purple-dotted-dashed, $\kappa = 100$) follows the attitude adjustment
(blue-dashed, $\kappa = 100$) curve at early times, later shifting
towards the network remodeling (red-solid) curve.  Guests thus find it
more advantageous to first adjust their attitudes, and then modify
their network connectivity.  Similarly, Fig.\,\ref{FIG:TEMP_EVOL}b
shows $\langle x_i^t \rangle_{\rm guest} \to 0^-$ as $t \to \infty$, 
following the curve where only attitude adjustment is allowed.  The convergence of
$\langle x_i^t \rangle_{\rm host} \to 0^{+}$ is slower because network
remodeling allows the many hosts to replace their relatively few
out-group connections with conspecifics.  Eventually however, both
guests and hosts converge towards integration, with $\langle x_i^t
\rangle_{\rm guest} \to 0^{-}$, $\langle x_i^t \rangle_{\rm host} \to
0^{+}$.  In contrast, $\langle U^t_{i} \rangle_{\rm guest}$ for the
full model with slow attitude adjustment (magenta-double-dotted-dashed
$\kappa = 1000$) follows the network remodeling (red-solid) curve at
all times. Here, guests find it more advantageous
to change their connectivity, preferentially creating links to other
guest nodes, rather than modify their attitudes towards host
communities.  Indeed attitudes converge to $\langle x_i^t \rangle_{\rm guest}
\to -0.34$ and $\langle x_i^t \rangle_{\rm host} \to 0.95$ as $t \to
\infty$, with no further attitude adjustment possible. 

This example illustrates the central role played by $\kappa$ in the
dynamics: low values of $\kappa$, indicating relatively short times
for attitude adjustment $\tau_{\rm g}, \tau_{\rm h}$, lead to
harmonious societies with $x_i \to 0$ for all nodes, while larger
values of $\kappa$, indicating longer times for attitude adjustment,
lead to segregated communities.

In Figs.\,\ref{FIG:TEMP_EVOL}c and\,\ref{FIG:TEMP_EVOL}d we show
$\langle U_i^t \rangle_{\rm guest}$ and $\langle x_i^t \rangle_{\rm
  guest}$ for a smaller immigrant population, $N_{\rm g} = 20$ and the
same parameters as in Figs.\,\ref{FIG:TEMP_EVOL}a
and\,\ref{FIG:TEMP_EVOL}b.  We observe the same qualitative increase
of utility in each of the five cases as discussed above.
Discrepancies with plots obtained for $N_{\rm g} = 200$ mainly emerge
when only attitude adjustment is allowed (blue-dashed $\kappa = 100$,
and green-dotted $\kappa = 1000$ curves). Here, the early increase of
utility is faster than for $N_{\rm g} = 200$, but steady state is
reached at a much slower rate. The overwhelming majority of hosts
drives guests to rapidly adjust their attitudes, increasing $\langle
U^t_{i} \rangle_{\rm guest}$ at short times. By the same token, the
host majority will not significantly change its attitude, so that
guests can further increase their utility only by remodeling their
connectivity.  Indeed, the corresponding curves in
Fig.\,\ref{FIG:TEMP_EVOL}d show guests rapidly converging to $\langle
x_i^t \rangle_{\rm guest} \to 0^-$ for all cases, while $\langle x_i^t
\rangle_{\rm host}$ does not. Note that as long as the network is
initially connected and no isolated clusters exist, when only attitude
adjustment is allowed, $\langle x_i^t \rangle_{\rm host} \to 0^+$ as 
$t \to \infty$, although the process may be slow. For $N_{\rm g} = 20$, 
due to the low number of guests, there is a higher probability than for 
$N_{\rm g} = 200$ of initiating the model with isolated host-only clusters. For
these clusters, if only attitude adjustment is allowed, attitudes will
stay quenched at $\langle x_i^t \rangle_{\rm host} \to 1$.  As a
result, the overall $\langle x_i^t \rangle_{\rm host}$ will converge
towards a non zero value.

%
\begin{figure}[t]
  \begin{center}
      \includegraphics[width=5.0in]{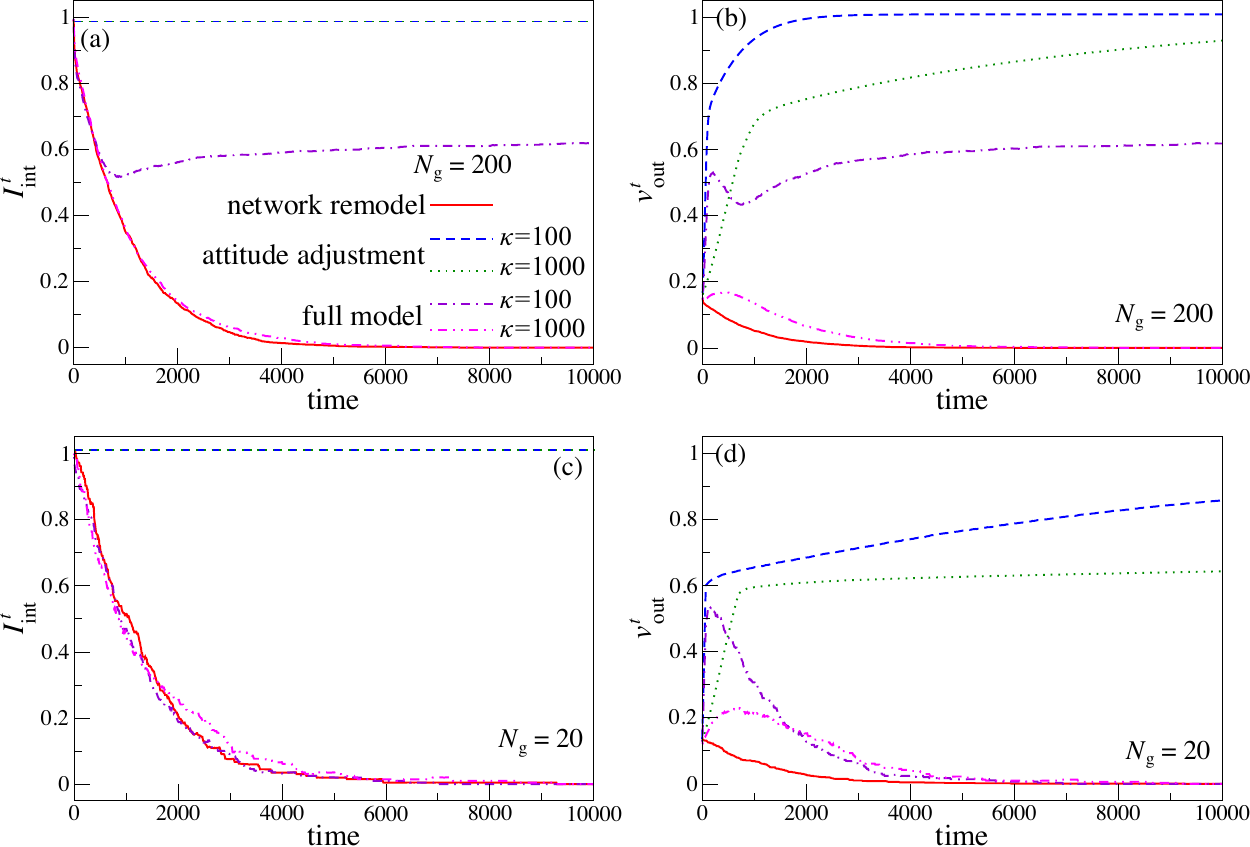}
\end{center}
  \caption{Dynamics of the integration index $I^t_{\rm int}$ in panels (a) and
    (c) and of the out-group reward fraction $v^t_{\rm out}$ in panels
    (b) and (d).  Parameters and initial conditions are the same as in
    Fig.\,\ref{FIG:TEMP_EVOL}.  (a, b) Large migrant population
    $N_{\rm g} = 200$.  Here, $I_{\rm int}^t \to 0$ and 
    $v_{\rm out}^t \to 0$ at long times when only network remodeling 
    is allowed, and nodes seek
    links with conspecifics.  If only attitude adjustment is allowed,
    $I_{\rm int}^t$ remains fixed due to the quenched network
    connectivity, while $v_{\rm out}^t$ increases as guests and hosts
    adopt more cooperative attitudes. For the full model, slow
    attitude changes ($\kappa = 1000$) lead to segregation and 
    $I_{\rm int}^t \to 0$, $v_{\rm int}^t \to 0$ as $t \to \infty$.  
    Fast attitude changes
    ($\kappa = 100$) lead to non-zero values of $I_{\rm int}^t$ and
    $v_{\rm out}^t$, indicating a more cooperative society.  (c, d)
    Small migrant population $N_{\rm g} = 20$. Results are similar to
    the previous case except for the full model where 
   $I_{\rm int}^t  \to 0$, $v_{\rm out}^t \to 0$ 
    as $t \to \infty$ for both $\kappa = 1000$ and $\kappa
    = 100$. For low values of $N_{\rm g}$ segregation arises under
    both fast and slow attitude changes.
    \label{FIG:TEMP_INT}}
\end{figure}

In the case of the full model (purple-dotted-dashed, $\kappa = 100$
and magenta-double-dotted-dashed $\kappa = 1000$) we see a similar
trend $\langle x_i^t \rangle_{\rm guest} \to 0^-$, while $\langle
x_i^t \rangle_{\rm host} \to 0.93$ for $\kappa = 100$, and $\langle
x_i^t \rangle_{\rm host} \to 0.99$ for $\kappa = 1000$ as $t \to
\infty$. Segregated host communities arise, with the numerically
lesser guests adapting to the majority.

\subsection{Quantifying outcomes of integration}
\label{SEC:TEMPORAL2}

The above results lead us to seek measures to better understand the
topology of the network as a function of time, specifically from the
guest standpoint.  To this end, we introduce an integration 
index
$I^t_{\rm int}$ 
as the relative number
of out-group connections of a guest node, averaged over all nodes, and
scaled by the host population fraction

\begin{equation}
\label{integrate}
  I^t_{\rm int} \equiv  \frac{N}{N_{\rm h}} \left< \frac{m_{i,{\rm out}}^{t}}{m^t_i} \right>_{\rm guest}.
\end{equation}

\noindent
Here,
$N_{\rm h} / N$ is the time independent host population fraction and 
$m_{i,{\rm out}}^t$ is
the number of out-group connections; 
the ratio $m_{i,{\rm out}}^t / m^t_i$
is averaged over all guest nodes. A guest-only enclave
for which $m_{i,\rm out}^t = 0$ leads to $I_{\rm int} =
0$. Conversely, in a uniformly mixed guest-host configuration, $m_{i, \rm
  out}^t / m^t_i$, should not be too dissimilar from the host
population fraction $N_{\rm h} / N$, leading to $I^t_{\rm int} \to 1$.
As defined, $0 \le I^t_{\rm int} \le N/N_{\rm h}$. At 
  $I^{\rm max}_{\rm int}= N/N_{\rm h} \geq 1$
guest nodes preferentially connect to hosts,
shunning other guest nodes. We refer to this outcome as
reverse segregation. 

While $I^t_{\rm int}$ measures the connectivity between guest and host
nodes, another relevant measure is the fraction of the reward
$u_{ij}^t$ that arises from cross-group interactions. This is
important, as guests connecting predominantly to host nodes may not
necessarily be an indicator of balanced socioeconomic growth.  For
example, even for large values of $I^{t}_{\rm int} \gtrsim 1$ hosts
may share large rewards among themselves but very little with guests,
representing a two-track society where guests, although connected, are
not part of the mainstream socioeconomic activity.

In a perfect scenario, guests and hosts form an all-connected network,
with $N_{\rm g} N_{\rm h}$ out-group, host-guest connections among the
total $N (N-1)/ 2$ edges.  If the reward is distributed equally among
all edges, the ratio of out-group connections is given by $2 N_{\rm g}
N_{\rm h} / N (N-1)$.  We thus define an out-group reward fraction
$v^t_{\rm out}$ as follows

\begin{equation}
   v^t_{\rm out} \equiv 
       \frac{\displaystyle \sum_{\substack{i \in {\rm guests} \\j \in {\rm hosts}}}
           u_{ij}^{t}} 
          {\displaystyle \sum_{\substack{i \in {\rm all\,nodes} \\ (j \ne i) \in {\rm all\,nodes}}}
           u_{ij}^{t} / 2} 
           \cdot
         \frac{N (N-1)}{2 N_{\rm g} N_{\rm h}}.
  \label{EQ:vout}
\end{equation}

\noindent
The first term on the right-hand side is the fraction of reward shared
between guests and hosts with respect to the total. We then
renormalize this quantity by the ratio $2 N_{\rm g} N_{\rm h} / N
(N-1)$ derived above for the perfectly mixed scenario.  As a result,
$v_{\rm out}^t = 1$ indicates a connected network with no isolated
clusters and with rewards equally spread among all nodes. Instead,
$v_{\rm out}^t = 0$ points to complete segregation, where no
socioeconomic reward comes from cross-group activities. Note that
$v^t_{\rm out}$ can exceed unity if the cross-group economy is more
flourishing than intra-group growth.

The dynamics of $I^t_{\rm int}$ and $v_{\rm out}^t$ under the same
parameter choices and mechanisms used to plot
Fig.\,\ref{FIG:TEMP_EVOL} are shown in Fig.\,\ref{FIG:TEMP_INT}.  We
first discuss the case of $N_{\rm g} = 200$, in
Figs.\,\ref{FIG:TEMP_INT}a and \ref{FIG:TEMP_INT}b.  If we allow only
for network remodeling (red-solid curves), the system will evolve
towards segregation ($I_{\rm int}^t \to 0$ in
Fig.\,\ref{FIG:TEMP_INT}a and $v_{\rm out}^t \to 0$ in
Fig.\,\ref{FIG:TEMP_INT}b).  Here, since attitudes cannot change,
nodes will maximize their utility through in-group connections
and by creating insular communities. In the blue-dashed and green-dotted
curves we deactivate network remodeling and only allow for attitude
adjustment, with $\kappa = 100, 1000$ respectively.  As can be seen
from Fig.\,\ref{FIG:TEMP_INT}a $I^t_{\rm int} \simeq 1$ at all times
since the random connections assigned at $t=0$ are fixed and guest and
host nodes remain well mixed in time.  Fig.\,\ref{FIG:TEMP_INT}b shows
that as cooperative attitudes emerge, cross-group rewards $v_{\rm
  out}^t$ increase.  In the case of fast attitude adjustment $\kappa =
100$ (blue-dashed curve), when nodes are completely cooperative,
$v^t_{\rm out} \to 1$ as $t \to \infty$, while in the case of slow attitude adjustment
$\kappa = 1000$ (green-dotted curve) convergence to $v^t_{\rm out} \to
1$ is slower.
 
Results for the full model reveal the subtle interplay between network
remodeling and attitude adjustment.  
At early times $I_{\rm int}^t$ follows the network remodeling case only 
(red-solid curve) for both $\kappa = 100$ and $\kappa = 1000$. In both 
scenarios guests progressively severe
their ties to hosts, due to their low utility.
At the same time, attitude adjustment increases cooperativity on the
given initial connections and $v^t_{\rm out}$ temporarily increases.
Eventually ineffective cross-group connections are completely
eliminated under slow attitude adjustment (magenta-double-dotted-dashed,
$\kappa = 1000$) where $I^t_{\rm int} \to 0$ and 
$v_{\rm out}^t \to 0$ as $t \to \infty$.  
Under fast attitude adjustment (purple-dot dashed, $\kappa = 100$)
instead cross-group connections contribute to the utility, so that
$I_{\rm int}^t \to 0.6$ and $v_{\rm out}^t \to 0.6$ as $t \to \infty$.
Note that $I_{\rm int}^t$ and $v_{\rm out}^t$ converge to the same
value as $t \to \infty$ since $\vert x_i^t - x_j^t \vert \to 0$ for 
both in-group and out-group connections. As a result, the distribution 
of rewards directly reflects the fraction of cross-group connections.

Taken together with results shown in Fig.\,\ref{FIG:TEMP_EVOL}a and
\ref{FIG:TEMP_EVOL}b, the above dynamics confirm the crucial role
played by $\kappa$, the attitude adjustment timescale, in determining
societal outcomes.  For the chosen parameters and when the full model
is considered, more rapid attitude adjustment ($\kappa=100$) leads to
a more integrated society with $I_{\rm int}^t $, $v_{\rm out}^t$
reaching non-zero values as $t \to \infty$, and with $\langle x_i^t
\rangle_{\rm guest} \to 0^{-}$ and $\langle x_i^t \rangle_{\rm host}
\to 0^+$.  All these are hallmarks of a well-mixed, functional
society, where guests and hosts share links, their socioeconomic
progress is intertwined, and groups are not hostile to each other.  On
the other hand, slower attitude adjustment ($\kappa=1000$) leads to a
segregated society, where $I_{\rm int}^t \to 0$, $v_{\rm out}^t \to
0$, and where $\langle x_i^t \rangle_{\rm guest}$ and $\langle x_i^t
\rangle_{\rm host}$ converge to non zero values as $t \to \infty$. In
this case, there are no links connecting guests and nodes, there is no
shared socioeconomic interest, and groups are hostile to each
other. Society is fragmented and parallel societies have emerged. Note
that these two opposite outcomes emerge from the same set of parameters,
with the exception of $\kappa$.

Because of their superior number, it is the attitudes of hosts in
particular that play a fundamental role in determining whether a
society is segregated or not. This is consistent with findings from
several surveys and societal observations \cite{KOO10,PRI14}.  Recall
that our initial conditions were set at $x_{i,\rm host} = 1$, the most
inhospitable.  Figs.\,\ref{FIG:TEMP_INT}a and \ref{FIG:TEMP_INT}b show
that this hostile environment drives the immigrant population towards
segregation, unless attitudes can easily change, {\textit {i.e.}} for
small $\kappa$.

Results for $N_{\rm g} = 20$ confirm the above scenario, with a small
difference.  Here, $I_{\rm int}^t \to 0$, $v_{\rm out}^t \to 0$ for
both values of $\kappa=100, 1000$ as $t \to \infty$,
while $\langle x_i^t \rangle_{\rm guest} \to 0^{-}$ and $\langle x_i^t
\rangle_{\rm host}$ converge to values that deviate only slightly
from unity. In this case, the very few guests must initially interact
with the many hosts and their attitude will become cooperative. Hosts
on the other hand will not necessarily link to guests, and due to
their numerical superiority can remain hostile towards them.  Over
time, separated enclaves of hosts and guests will emerge, with guests
keeping their cooperative attitude, but in isolation from hosts, while
hosts will largely remain in the same state as at the onset of the
adaptation process.  In this case, in order for a more cooperative
society to emerge the value of $\kappa$ must be even smaller. We have
verified this numerically, finding that for $N_{\rm g} = 20$, $\kappa
\lesssim 40$ in order for a more integrated society to emerge.

\subsection{Initially hostile host attitudes drive immigrants
into enclaves}
\label{SEC:TEMPORAL3}

%
\begin{figure}[t]
  \begin{center}
      \includegraphics[width=5.0in]{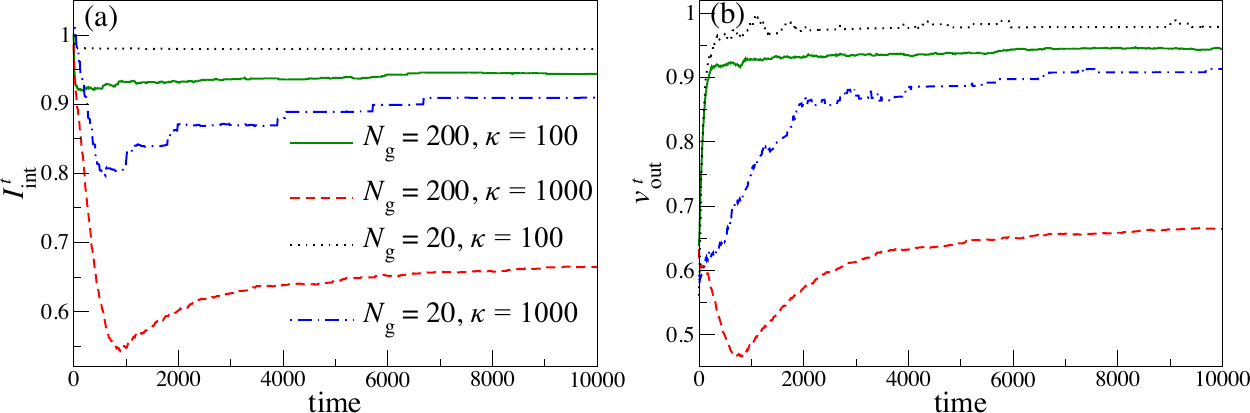}
\end{center}
  \caption{Dynamics of the integration index $I^t_{\rm out}$ in panel (a) and
    of the out-group reward fraction $v^t_{\rm out}$ in panel (b) for
    initially cooperative hosts.  Parameters are the same as for the
    full model in Fig.\,\ref{FIG:TEMP_EVOL}, with initially
    cooperative hosts and uncooperative guests at $x_{i,{\rm host}}^0
    = 0^+$ and $x_{i,{\rm guest}}^0 = -1$.  (a) $I_{\rm int}^t$
    decreases at the onset, eventually rising towards integration,
    where $I_{\rm int}^t \to 1$ as $t \to \infty$.  
    The initial decrease is more pronounced
    for slow attitude adjustment ($\kappa = 1000$) and for larger
    guest populations ($N_{\rm g} = 200$) as described in the text.
    (b) $v_{\rm out}^t$ increases over long times as attitude
    adjustment allows for more cooperation between guests and
    hosts. Under slow attitude adjustment ($\kappa = 1000$) and large
    guest populations ($N_{\rm g} = 200$), $v_{\rm out}^t$ decreases
    at the onset, with players seeking in-group connections. As guests
    and hosts become more cooperative $v_{\rm out}^t$ increases.}
  \label{FIG:TEMP_HOST}
\end{figure}
%

The importance of initial attitudes is further examined in
Fig.\,\ref{FIG:TEMP_HOST}, where at $t=0$ hosts are extremely
hospitable and $x_{i, {\rm host}}^{0} = 0$.  Initial guest attitudes
remain uncooperative at $x_{i, {\rm guest}}^0 = -1$. All other
parameters are set as in Figs.\,\ref{FIG:TEMP_EVOL} and
\ref{FIG:TEMP_INT}.  Curves in Fig.\,\ref{FIG:TEMP_HOST}a and
\ref{FIG:TEMP_HOST}b arise from the full model and should be compared
to their counterparts in Fig.\,\ref{FIG:TEMP_INT}a and
\ref{FIG:TEMP_INT}c.

In Fig.\,\ref{FIG:TEMP_HOST}a we plot $I^t_{\rm int}$. As can be seen,
guests and hosts are no longer completely segregated. At early times,
$I_{\rm int}^t$ decreases due to network remodeling, however at
intermediate times, guests become more cooperative so that
$x_{i,{\rm{guest}}}^t \to 0^{-}$ and $I_{\rm int}^t \to 1$ for long
times.  The early decrease of $I_{\rm int}^t$ is more significant for
$\kappa = 1000$, since slow attitude adjustment leads to ineffective
cross-group links and network remodeling will induce segregation. The
decrease of $I_{\rm int}^t$ is also relatively more significant for
$N_{\rm g} = 200$ than for $N_{\rm g} = 20$ under the same value of
$\kappa$. This is because a larger guest population, and a larger
$\tau_{\rm g} = \kappa N/N_{h}$ will more slowly evolve its initially
hostile attitudes, allowing for segregation to cut cross-group,
ineffective connections.  In Fig.\,\ref{FIG:TEMP_HOST}b, we plot
$v_{\rm out}^t$ which increases at early times in all cases except for
$N_{\rm g} =200$ and under slow adjustment $\kappa = 1000$. This is
due to network remodeling. As discussed above, the slow attitude
adjustment prompts nodes to seek in-group connections at early times;
the guest population is large enough to allow for this leading to
segregation with $I_{\rm int}^t \simeq 0.5$ and an initially
decreasing $v_{\rm out}^t$ for the red-dashed curve.  Due to the
cooperative attitude of hosts however, guest eventually change their
attitudes so that $x_{i,{\rm{guest}}}^t \to 0^{-}$, and $I_{\rm int}^t$
and $v_{\rm out}^t$ increase.
  
One interesting finding is that when initial host attitudes are
hostile, as shown in Figs.\,\ref{FIG:TEMP_INT}c and
\ref{FIG:TEMP_INT}d, a larger guest population 
more effectively 
drives host attitudes
towards cooperation.  In contrast, when initial host attitudes are
hospitable, as shown in Fig.\,\ref{FIG:TEMP_HOST}a, a larger guest
population results in less integration. In this case, the larger guest
population is more resistant to attitude changes, and segregation may
more easily emerge.

\subsection{Higher initial connectivity facilitates better
integration}
\label{SEC:TEMPORAL4}

%
\begin{figure}[t]
  \begin{center}
      \includegraphics[width=5.0in]{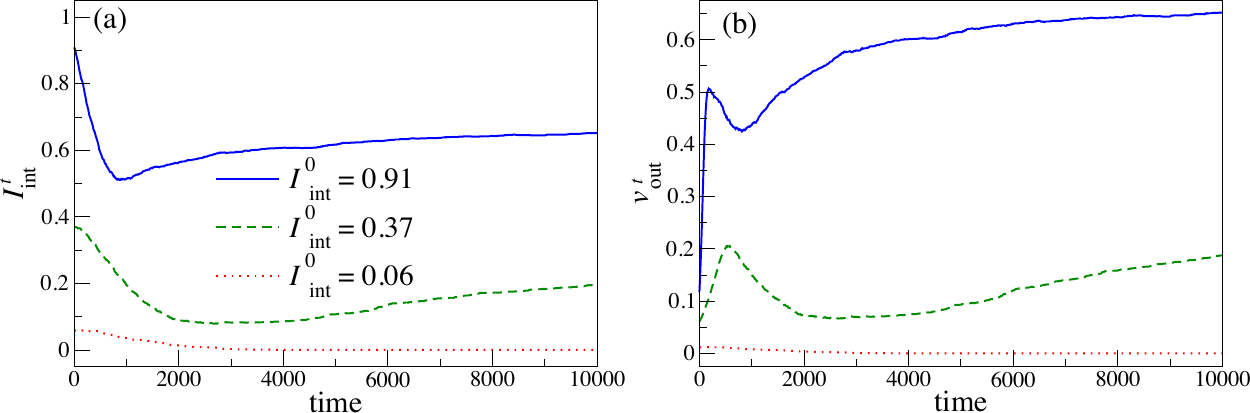}
\end{center}
  \caption{Dynamics of the integration index $I^t_{\rm out}$ in panel (a) and
    of the out-group reward fraction $v^t_{\rm out}$ in panel (b) under
    different initial random connectivities.  Parameters are the same
    as in Fig\,\ref{FIG:TEMP_EVOL} with initial hostile attitudes
    $x_{i,{\rm host}}^0 = 1$ and $x_{i,{\rm guest}}^0 = -1$.  In the
    blue-solid curve $I_{\rm int}^0 = 0.91$; in the green-dashed curve
    $I_{\rm int}^0 = 0.37$; in the red-dotted curve $I_{\rm int}^0 =
    0.06$.  (a) For all three cases, $I_{\rm int}^t$ decreases from
    the initial values, but only the initially poorly connected case
    of $I_{\rm int}^0 = 0.06$ leads to full segregation, indicated by $I_{\rm
      int}^t \to 0$ as $t \to \infty$. For the other two cases, 
      $I_{\rm int}^t \to 1$. (b) For all three cases $v_{\rm out}^t$
    increases at the onset due to attitude adjustment, and later
    decreases due to network remodeling.  Only $I_{\rm int}^0 = 0.06$
    leads to long-time $v_{\rm out}^t \to 0$: as guest-host connections are
    severed, no socioeconomic utility can be shared.  For the other
    two cases, $v_{\rm out}^t$ increases at long times, suggesting
    increasing rewards through cross-group connections.}
  \label{FIG:TEMP_CONN}
\end{figure}
%

The initial social connections assigned to 
migrants upon arrival may
affect integration outcomes. As discussed in Section \ref{SEC:IC}, one
ideal scenario is that of welcoming programs that provide guests
with prearranged social connections to hosts ($I_{\rm int}^0 = 1$), another is that of
completely isolated guests arriving in an already connected native
society ($I_{\rm int}^0 = 0$).
In previous sections we only implemented these two extremes,
perfect connectivity or total isolation. In this section we will
consider 
more realistic, intermediate
levels of initial guest connectivity.

Figure.\,\ref{FIG:TEMP_CONN} 
illustrates the effects of three initial
configurations: well connected guests, $I_{\rm int}^0 = 0.91$
(blue-solid curve), intermediately connected guests, $I_{\rm int}^0 =
0.37$ (green-dashed curve), and poorly connected guests $I_{\rm int}^0
= 0.06$ (red-dotted curve). Initial attitudes are uncooperative,
$x_{i,{\rm host}}^0 = 1$ and $x_{i, {\rm guest}}^0 = -1$. All model
parameters are the same as in Fig.\,\ref{FIG:TEMP_EVOL} with $N_{\rm
  g} = 200$ and $\kappa = 100$.

The time evolution of $I_{\rm int}^t$ for all cases is shown in
Fig.\,\ref{FIG:TEMP_CONN}a. Here, $I_{\rm int}^t$ decreases at early
times until guests and hosts begin adopting more cooperative
attitudes. For the initially well connected case (blue-solid curve),
$I_{\rm int}^t$ drops to $I_{\rm int} \simeq 0.5$ before the trend is
reversed at $t \sim 1000$.  For the initially intermediately connected
case (green-dashed curve), the decreasing trend is not reversed until
$t \simeq 2500$ when $I_{\rm int}^t \simeq 0.1$.  Finally, for the
initially poorly connected case (red-dotted curve) attitude adjustment
cannot give rise to cooperation before $I_{\rm int}^t \to 0$ and host
and guest communities are fully segregated.  Mirroring trends are seen
in Fig.\,\ref{FIG:TEMP_CONN}b where we plot the out-group reward
fraction $v_{\rm out}^t$.  When
guests are poorly connected at the onset (red-dotted curve), few links
exists through which attitudes can change, guests become progressively
segregated, and very little socioeconomic activity is shared. Hence,
$v_{\rm out}^t \to 0$ throughout.  For the other two cases when there
is more initial connectivity at the onset $v_{\rm out}^t$ increases at
early times (blue-solid and red-dotted curves) as guests adopt
cooperative attitudes ($x_{i, {\rm guest}}^t \to 0^-$) through these
initial guest-host connections.  Later, network remodeling causes
$v_{\rm out}^t$ to decline as cross-group connections are replaced
with in-group ones.  At longer times, host attitudes also evolve
towards cooperation ($x_{i, {\rm host}} ^t \to 0^+$) from residual
guest-host connections. Here, network remodeling no longer favors
in-group connections, and $v_{\rm out}^t$ increases once more.

Although these results point to the importance of an initial network
of connections for immigrants, in reality very few of them will have a
support system upon arrival. Many host countries may not have adequate
resources or programs to foster such contact, and host and guest
communities may view each other with suspicion.  In the rest of this
paper we attempt to identify best practices leading to integration,
and look at how results vary depending on model parameters. We will
consider a realistic, worst case initial condition: that of an
initially equilibrated host community and a totally isolated guest
cohort, as outlined in Section \ref{SEC:IC}.

\subsection{Dependence on parameters of
cross-group reward, attitude adjustment rate, and sensitivity
to attitude difference}

We now
study how results from the model defined in
Eqs.\,\ref{EQ:UTILITY}--\ref{EQ:CONFORMATION} depend on its main
parameters $\alpha$, $A_{\rm in}$, $A_{\rm out}$, $\sigma$, and
$\kappa$.  In earlier sections, we set $A_{\rm in} = A_{\rm out} = A$
and determined analytically that the utility reaches a maximum
$U_i^{\rm max} = \alpha A [\ln (\alpha A) - 1]$ if each node has
$m_{\rm opt} = \alpha \ln (A \alpha)$ links.  We have also verified
this numerically for several $\alpha, A$ parameter choices. Note that
setting $\alpha \lesssim A^{-1/2}$ leads to $m_{\rm opt} \lesssim 1$
indicating a network with no links, which we have verified
numerically.  We also briefly discussed how $\kappa$ affects the
dynamics by comparing results from high ($\kappa = 1000$) and a low
($\kappa = 100$) regimes.  Here we will conduct a more thorough
investigation of the relevant parameters.

\begin{figure}[t]
  \begin{center}
      \includegraphics[width=5.0in]{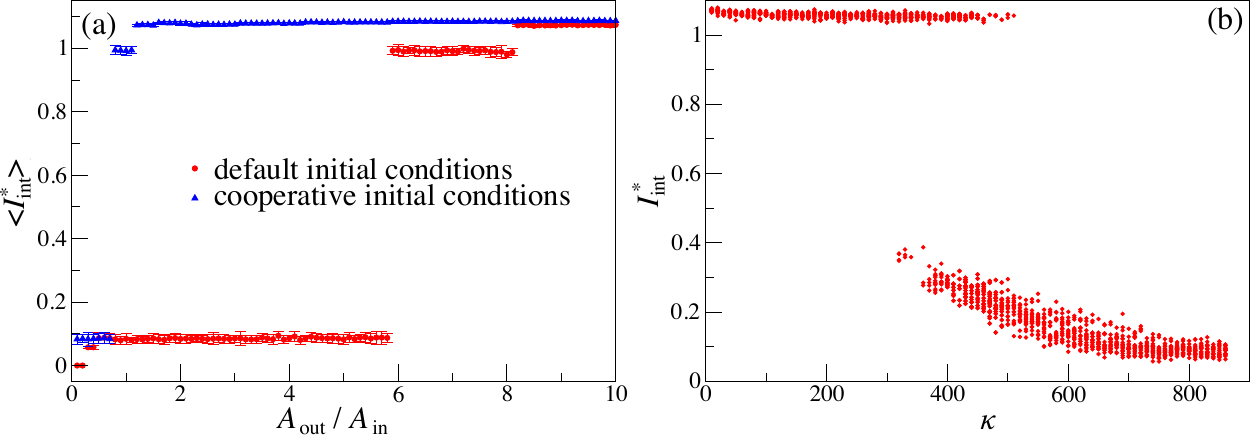}
\end{center}
  \caption{Integration index at steady state. In panel (a) $\langle I^*_{\rm
      int} \rangle$ is averaged over 20 realizations and plotted as a
    function of $A_{\rm out} / A_{\rm in}$ with $\kappa = \infty$. The
    bar indicates the variance.  In panel (b) single representations
    $I^*_{\rm int}$ are shown as a function of $\kappa$ with $A_{\rm
      out} / A_{\rm in} = 2$.  Other parameters are set at $\alpha =
    3$ and $\sigma = 1$, with $N_{\rm h} = 1800$ and $N_{\rm g} =
    200$.  In both panels red solid circles represent initially
    unconnected, hostile hosts and guests, $x_{i,{\rm host}}^0 = 1$,
    $x_{i,{\rm guest}}^0 = -1$; blue triangles correspond to fully
    cooperative initial conditions $x_{i,{\rm host}}^0 = x_{i,{\rm
        guest}}^0 =0$.  When the ratio $A_{\rm out} / A_{\rm in}$
    increases, the long-time state of the network changes from
    segregation to uniform mixture, and finally to reversed
    segregation. The transition for the default initial conditions
    occurs at larger $A_{\rm out} / A_{\rm in}$ ratios, compared to
    the cooperative initial conditions, as the former require higher
    compensation from out-group connections to overlook the hostile
    attitudes between guests and hosts.  In panel (b) each data point
    corresponds to one realization. Increasing attitude adjustment
    time scale $\kappa$ leads to increased likelihood of
    segregation. A bimodal regime emerges for intermediate $\kappa$.}
  \label{FIG:A_RATIO}
\end{figure}

First we examine a scenario where $A_{\rm out} \ne A_{\rm in}$ and the
effects of varying $A_{\rm out} / A_{\rm in}$ while keeping other
parameters fixed.  In Fig.\,\ref{FIG:A_RATIO}a we plot the
steady-state integration index $\langle I_{\rm int}^* \rangle$ as a
function of $A_{\rm out} / A_{\rm in}$, with $A_{\rm in} = 10$,
$\kappa \to \infty$, $\alpha = 3$, and $\sigma = 1$ for $N_{\rm g} =
200$ guests and a total population of $N = 2000$ nodes, corresponding
to $N/N_{\rm h} = 1.11$. Note that setting $\kappa \to \infty$ is
equivalent to activating network remodeling only, since the timescale
for attitude change diverges, hence attitudes $x_i^{t}$ will remain
fixed at their initial values throughout the entire course of the
dynamics.  We also use two different initial conditions of total guest
isolation but different initial attitudes.  The blue-solid triangles
represent initially cooperative populations with $x_{i,{\rm host}}^0 =
x_{i,{\rm guest}}^0 = 0$, while the red-solid circles represent
initially hostile populations with $x_{i,{\rm host}}^0 = 1$, and
$x_{i,{\rm guest}}^0 = -1$.  In both cases, guests have no connections
at $t=0$.

%

Each data point and relative error bar in Fig.\,\ref{FIG:A_RATIO}a
represents the mean and variance over $20$ realizations,
respectively. For the cooperative case (blue-solid triangles) as long
as $A_{\rm out} / A_{\rm in} \lesssim 1$ in-group connections yield
higher rewards and are preferred; hence the two populations are almost
completely segregated and $\langle I_{\rm int}^* \rangle \to
0.1$. Conversely, when $A_{\rm out} / A_{\rm in} \gtrsim 1$ out-group
connections are preferred, and $\langle I_{\rm int}^* \rangle \to N /
N_{\rm h} = 1.11$, indicating reverse segregation.  When $A_{\rm out}
/ A_{\rm in} \simeq 1$ out-group and in-group connections are
equivalent in terms of their socioeconomic weight and integration is
observed at $\langle I_{\rm int}^* \rangle \to 1$.  Note the sharp
transitions between regimes.  The progression segregation $\to$
integration $\to$ reverse segregation as a function of $A_{\rm out} /
A_{\rm in}$ also appears for the uncooperative conditions (red-solid
circles).  However, in this case transitions are shifted towards the
right, indicating that out-group connections must yield higher
socioeconomic gain to promote integration (or reverse segregation) in
order to overcome the initial hostility among players.  Here,
segregation persists until $A_{\rm out} / A_{\rm in} \lesssim 6$ for
which $\langle I_{\rm int}^* \rangle \lesssim 0.1$, full integration
$\langle I_{\rm int}^* \rangle \to 1$ arises for $6 \lesssim A_{\rm
  out} / A_{\rm in} \lesssim 8$ and reverse segregation at $\langle
I_{\rm int}^* \rangle \to N/N_{\rm h} = 1.11$ appears only for $A_{\rm
  out} / A_{\rm in} \gtrsim 8$.  Note that in both cases since attitudes
are fixed, rewards are given by $u_{ij} = A_{\rm out} \, e^{-2}$ if
through out-group connections, and by $u_{ij} = A_{\rm in}$ if through
in-group ones. The two will be the same for $A_{\rm out} / A_{\rm in}
= e^2 = 7.39$.
 
These results indicate that to promote integration, cross-group
connections must generate higher rewards than in-group ones.  This may
be realized, for example, if the immigrant population possesses skill
sets that complement those of the host population.  Since no attitude
adjustment is allowed in the dynamics, Fig.\,\ref{FIG:A_RATIO}a
suggests that integration may occur even if groups maintain their
hostility towards each other as long as the socioeconomic rewards are
large enough, as seen for the uncooperative case (red-solid
circles). Finally, note that the same parameter sets yields very
different results for a wide range of $A_{\rm out} / A_{\rm in}$
values, as can be seen by the bimodal values of $\langle I_{\rm int}^*
\rangle$ in Fig.\,\ref{FIG:A_RATIO}a and underlying the role of
initial conditions in determining integration or segregation.

We examine the effects of varying $\kappa$ in
Fig.\,\ref{FIG:A_RATIO}b. Here, we use the same parameters as in
Fig.\,\ref{FIG:A_RATIO}a with $A_{\rm in} = 10$, $A_{\rm out} = 20$,
$\alpha = 3$, and $\sigma = 1$.  The ratio $A_{\rm out} / A_{\rm in} =
2$ provides modest incentives for guests and hosts to collaborate.  We
consider initially hostile guests and hosts at $x_{i,{\rm host}}^0 =
1$ and $x_{i,{\rm guest}}^0 = -1$, and omit fully cooperative initial
conditions $x_{i,{\rm host}}^0 =x_{i,{\rm guest}}^0 = 0$ since, in
this case, changes to $\kappa$ will not alter the dynamics. Each red
solid circle in Fig.\,\ref{FIG:A_RATIO}b is the result of a single
simulation; for each value of $\kappa$ simulations are repeated $20$
times.  The dot-plot shows that if attitude adjustment is sufficiently
fast ($\kappa \lesssim 300$) reverse segregation arises and $I^*_{\rm
  int} \simeq N/N_{\rm h} = 1.11$; guests and hosts adopt cooperative
attitudes before segregation can arise. For very slow attitude
adjustments ($\kappa \gtrsim 550$), almost complete segregation as
$I^*_{\rm int} \to 0.1$ is the only outcome. A bimodal regime instead
arises for intermediate values of $300 \lesssim \kappa \lesssim 550$
where segregation and reverse segregation are both likely.  The
bimodal feature of Fig.\,\ref{FIG:A_RATIO}b is indicative of the
different timescales between the two competing mechanisms of network
remodeling and attitude adjustment.  If attitude adjustment is fast
compared to network remodeling (low $\kappa$) guests will quickly
adopt cooperative attitudes, and guest-only enclaves will not be
formed. Conversely, if attitude adjustment is slow compared to network
remodeling (large $\kappa$) guest-only enclaves will form hindering
cooperativity. In between these limits, is a regime where the
timescales of network remodeling and attitude adjustment are
comparable, and the outcomes stochastic.

%
\begin{figure}[t]
  \begin{center}
      \includegraphics[width=6.5in]{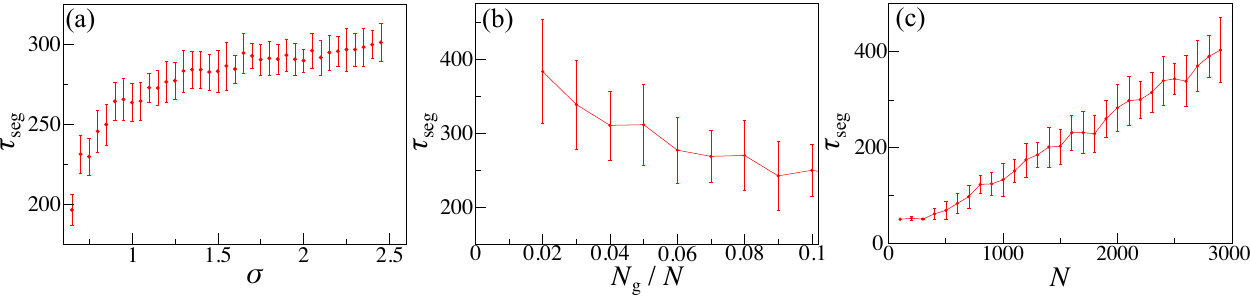}
\end{center}
  \caption{Time $\tau_{\rm seg}$ to reach $\langle I_{\rm int}^*\rangle =
    0.1$, where 90$\%$ of guest nodes are segregated as a function of
    (a) the sensitivity to the reward function $\sigma$, (b) the
    relative guest population $N_{\rm g}/N$ and (c) the total
    population $N$ assuming $N_{\rm g} = 0.1 N$.  Other parameters are
    set to $\alpha = 3$, $A_{\rm in} = A_{\rm out} = 10$, $\kappa =
    600$ in all panels. In panel (a) $N_{\rm g} = 200$, $N=2000$; in
    panel (b) $\sigma = 1$ and $N=2000$; in panel (c) $\sigma = 1$.
    In all three cases, guests and hosts are initially unconnected and
    hostile to each other, $x_{i,{\rm host}}^0 = 1$ and $x_{i,{\rm
        guest}}^0 = -1$.  Each data point and its error bar represent
    the mean and the variance over $20$ simulations.  In panel (a)
    increasing $\sigma$ allows for more tolerance to attitude
    differences, increasing the time to segregation. In panel (b) the
    higher guest population ratio leads to faster segregation as
    guests are more likely to establish in-group connections, 
    forming guest only enclaves. In panel (c) the time to
    segregation increases with the overall population, for a constant
    $10\%$ guest population.
  \label{FIG:SIGMA}}
\end{figure}

The last parameter we examine here is $\sigma$, which regulates the
sensitivity of the reward function $u_{ij}^t$ to attitude differences
$\vert x_i^t - x_j^t \vert$ in Eq.\,\ref{EQ:UTILITY}.  Note that
$\sigma \to \infty$ renders $u_{ij}^t$ independent of $\vert x_i^t -
x_j^t \vert$.  Finite values of $\sigma$, however large, do not
determine whether in-group or out-group connections are
preferred. This parameter thus will only affect the timescale of the
dynamics.  In particular, since larger values of $\sigma$ attenuate
the sensitivity of $u_{ij}^t$ to $\vert x_i^t - x_j^t \vert$ we expect
larger values of $\sigma$ to also be associated with slower dynamics.
We have verified this by considering the time to reach $90\%$
segregation, defined as $I^t_{\rm int} = 0.1$, as a function of $\sigma$
and for a variety of parameter choices.  In Fig.\,\ref{FIG:SIGMA}a we
plot the time to segregation, denoted by $\tau_{\rm seg}(\sigma)$, for
the particular case of $\alpha = 3$, $A_{\rm in} = A_{\rm out} = 10$,
and $\kappa = 600$, with initially hostile populations $x_{i,{\rm
    host}}^0 = 1$ and $x_{i,{\rm guest}}^0 = -1$ and no initial link
between hosts and guests. As can be seen, $\tau_{\rm seg} (\sigma)$
increases with $\sigma$.  This result suggests that 
decreasing
 the
sensitivity to attitude differences, particularly between guests and
hosts, results in longer times to full segregation. This larger time
window between migrant arrival and full segregation may provide better
opportunities to implement interim policies that promote cooperation.

\subsection{High immigrant ratios and small native populations promote segregation}

\noindent
In this section we examine the effects of migrant population sizes
compared to that of the native community. We are particularly
interested in the uncooperative, segregated case and examine how the
time to segregation $\tau_{\rm seg}$ depends on the fraction of
guests.  Under parameters and conditions that favor segregation, 
we expect larger guest populations will more quickly evolve to the 
uncooperative steady state.  We thus consider a scenario where at
steady state guests segregate, resulting in $x^t_{i,\rm{guest}} \to
-1$, $x^t_{i,\rm{host}} \to 1$, $I_{\rm int}^t \to 0$ as $t \to \infty$.  We then keep
all parameters fixed, including the total population $N$, and modify
only $N_{\rm g}$ to study $\tau_{\rm seg}$ as a function of the
$N_{\rm g}/N$ ratio. In Fig.\,\ref{FIG:SIGMA}b we show $\tau_{\rm
  seg}(N_{\rm g}/N)$ for the representative case of $\alpha= 3$,
$A_{\rm in} = A_{\rm out} = 10$, $\sigma = 1$, and $\kappa = 600$.
Initial conditions are initially hostile populations $x_{i,{\rm
    host}}^0 = 1$ and $x_{i,{\rm guest}}^0 = -1$ and no initial link
between hosts and guests. As can be seen, $\tau_{\rm seg}(N_{\rm
  g}/N)$ is a decreasing function of its argument, as expected.  Here,
attitude adjustment timescales $\kappa$ are not affected by $N_{\rm
  g}/N$, however a larger guest population makes in-group interactions
more likely under the dynamics specified in Section \ref{mech}. The
increased guest-guest pairing allows for uncooperative attitudes to be
maintained for longer times, lowering the utility reward from
cross-group interactions and hastening the severing of such links.
Numerically lower guest populations instead carry a higher likelihood
of interacting with hosts, fostering cooperative attitudes for longer
times, and allowing for socioeconomically advantageous cross-group
connections to emerge. Several sociological reports show that
conflicts between a majority $N_{\rm h}$ and a minority $N_{\rm g}$
population are less intense and frequent, if the majority population
greatly exceeds that of the minority, $N_{\rm h} \gg N_{\rm g}$
\cite{BAR04}.  We can conjecture that such conflicts arise when hosts
and guests are extremely polarized and segregated from each other, as
for the case illustrated above. Our results show that as $N_{\rm g}/
N$ increases segregation, and by proxy, the emergence of conflict
between the two groups increases as well, confirming these
sociological findings.

Finally, in Fig.\,\ref{FIG:SIGMA}c we plot $\tau_{\rm seg}(N)$ as a
function of the total population $N$ by fixing $N_{\rm g} = 0.1 N$.
All other parameters and initial conditions are the same as in
Fig.\,\ref{FIG:SIGMA}b.  As can be seen, larger $N$ populations lead
to longer times to segregation $\tau_{\rm seg} (N)$. This result
implies that the same fraction of migrants can be more easily
accommodated in larger communities.

%
\begin{figure}[t]
  \begin{center}
      \includegraphics[width=5.0in]{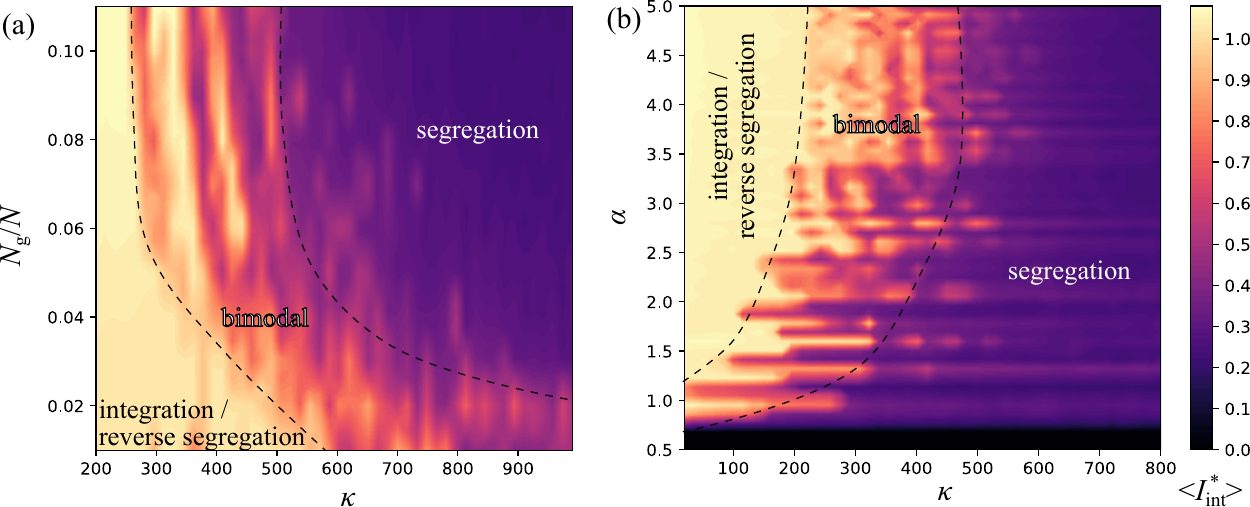}
\end{center}
  \caption{Integration index at steady state. $\langle I^*_{\rm
        int} \rangle$ is averaged over 10 realizations and plotted as
      a function of $\kappa$ and $N_{\rm g} / N$ with $\alpha = 3$ in
      panel (a), and as a function of $\kappa$ and $\alpha$ with
      $N_{\rm g} / N = 0.1$ in panel (b).  Other parameters are set at
      $A_{\rm in} = 10$, $A_{\rm out} = 20$, $\sigma = 1$, and $N =
      2000$.  In both panels guests and hosts are initially
      unconnected, with hostile attitudes, $x_{i,{\rm host}}^0 = 1$,
      $x_{i,{\rm guest}}^0 = -1$.  In panel (a), for smaller $N_{\rm
        g} / N$, the transition from segregation to integration 
      (or reverse segregation) occurs at larger $\kappa$. In panel (b) increasing $\alpha$
      causes the transition point to shift towards larger $\kappa$.}
  \label{FIG:KAPPA_MAP}
\end{figure}

\subsection{Transitioning from segregation to integration}

In this section we study the interplay between the two timescales, $\kappa$
and $\tau_{\rm seg}$, that determine whether or not guest-only enclaves
will form, starting from an initially hostile and unconnected mixture of guests and hosts.
In Fig.\,\ref{FIG:A_RATIO}b we showed that fast attitude adjustment
(small $\kappa$) prevents the formation of guest-only enclaves if 
incentives are in place to support cross-group collaborations ($A_{\rm out} / A_{\rm in} > 1$).
As shown in Fig.\,\ref{FIG:SIGMA}b, increasing the guest population ratio $N_{\rm g} / N$, 
shortens the time to segregation $\tau_{\rm seg}$ and facilitates the establishment of
guest-only enclaves.  

To study the interplay between $\kappa$ and $\tau_{\rm seg}$ 
 we plot $\langle I^*_{\rm int} \rangle$ in Fig.\,\ref{FIG:KAPPA_MAP}a as a function of
$\kappa$ and $N_{\rm g} / N$ for the representative case of $\alpha= 3$,
$A_{\rm in} = 10$, $A_{\rm out} = 20$, $\sigma = 1$, and $N=2000$.
The populations are initiated with hostile attitudes $x_{i,{\rm
    host}}^0 = 1$ and $x_{i,{\rm guest}}^0 = -1$, and no cross-group initial link. 
    As can be seen, decreasing $\kappa$ induces a
transition from segregation at $\langle I^*_{\rm int} \rangle \to 0$ for large $\kappa$, 
to integration at $\langle I^*_{\rm int} \rangle \to 1$, for small $\kappa$, or even reverse
segregation at $\langle I^*_{\rm int} \rangle \to N / N_{\rm h}$, for very small $\kappa$.
Transitions towards integration thus are favored in societies where attitudes towards the other
are less entrenched and where guests and hosts more readily adapt to each other.  
Fig.\,\ref{FIG:KAPPA_MAP}a also shows that transitions depend on the value of $N_{\rm g} / N$, 
and indirectly on $\tau_{\rm g}$:
larger values of $N_{\rm g} / N$ imply shorter transition $\kappa$ values.
This is because increases in $N_{\rm g} / N$, and consequently decreases in $\tau_{\rm seg}$,
correspond to less time for attitude adjustment to affect cross-group utility gains.
Larger percentages of migrants $N_{\rm g} / N$ imply that individual attitudes $\kappa$ must be
even more open to diversity if one is to observe the same target integration index
$\langle I^*_{\rm int} \rangle$. Note that in Fig.\,\ref{FIG:KAPPA_MAP}a we can also identify a
bimodal regime, where $\langle I^*_{\rm int} \rangle$ takes on values
between zero and $N / N_{\rm h}$ where final integration outcomes depend
on stochastic events.

Finally, in Fig.\,\ref{FIG:KAPPA_MAP}b, we study how the integration index $\langle I^*_{\rm int} \rangle$ 
depends on $\kappa$ and $\alpha$, the latter controlling the average number of connections
associated with each node. We fix $N_{\rm g} = 200$ and use the same parameter
values and initial conditions as in Fig.\,\ref{FIG:KAPPA_MAP}a. 
Increasing $\alpha$ corresponds to increasing the number of
connections per node. As can be seen in Fig.\,\ref{FIG:KAPPA_MAP}b 
the same progression seen in Fig.\,\ref{FIG:KAPPA_MAP}a 
of transitioning from segregation to integration can be seen 
upon lowering $\kappa$ for fixed $\alpha$. 
Increasing $\alpha$ leads these transition points to shift towards larger
values of $\kappa$, signifying that more connections per node 
allow for slower attitude adjustment to achieve the same integration value
$\langle I^*_{\rm int} \rangle$. Beyond $\alpha \gtrsim 3$ however,
the transition regime of $\kappa$ appears not to change appreciably, implying
little sensitivity of $\langle I^*_{\rm int} \rangle$ to the average number of connections per node.

\section{Discussion and Conclusions\label{SEC:DISCUSSION}}

As recent news reports and historical analysis attest, societal
dynamics after the influx of newcomers depends on many factors,
including the socioeconomic environment of the host country, the
adaptability of the immigrant population, the open-mindedness of
natives, and the degree of compatibility between guest and host
values.  Our model is based on the assumption that upon resettlement
immigrants have two primary goals: socioeconomic prosperity and social
acceptance.  Game-theoretic rules are used to model socioeconomic gains
through a utility function to be maximized, leading to network
remodeling. Attitude adjustment is instead driven by opinion dynamics
rules.  The two processes occur at different timescales: network
remodeling at a timescale of unity, and attitude adjustment at a
timescale of $\tau_{\rm g} = \kappa N/N_{\rm h} $ for guests.  Due to
their numerical superiority, hosts constitute a quasi-infinite bath: 
they greatly impact migrant dynamics, but their own characteristics change
only marginally and over very long timescales, given by $\tau_{\rm h}
= \kappa N/N_{\rm g} \gg \tau_{\rm g}$.

The interplay between the various timescales is shown across our analysis. 
For low values of $\kappa$ attitude adjustment is fast, cross-group
socioeconomic gains are robust and immigrants are less likely to form
segregated enclaves.  For large values of
$\kappa$, attitudes change very slowly, and the formation of isolated
guest niches becomes the most efficient way for guests to advance
their socioeconomic status.  Our results are consistent with findings
from public goods evolutionary game theory models where 
interactions among various social contexts, such as 
population diversity and cultural tolerance, lead to different
ratios between the timescales for strategy evolution and network structure
remodeling; such timescale difference
 determines whether cooperative patterns emerge \cite{SZO16,WAN18}. In
particular, cooperators will outweigh defectors if strategy evolution
is faster than network remodeling \cite{SAN06}.

The socioeconomic reward structure associated with guest-host
collaborations also plays an important role in determining societal
outcomes. As shown in Fig.\,\ref{FIG:A_RATIO}a larger out-group versus
in-group rewards, represented by the $A_{\rm out}/A_{\rm in}$ ratio,
are more conducive to integration.
Out-group rewards promote the willingness among a mixed
population to pursue conformity, which was identified as a key
psychological factor for cooperative patterns to emerge in
game theoretical models \cite{GAL15}.
 These results suggest that
segregation may be avoided if newcomers carry inherent advantages, for
example in the form of skill sets that are complementary to those of
the native population, or if governmental incentives are established
to promote cross-group interactions.  Fig.\,\ref{FIG:A_RATIO}a also
reveals the fundamental role of initial conditions.  If out-group
rewards are much larger than in-group ones, $A_{\rm out} \gg A_{\rm
  in}$, cooperation arises regardless of initial conditions. However,
if the two are comparable, Fig.\,\ref{FIG:A_RATIO}a shows that
integrated or segregated societies can emerge from the same parameter
set, and that whether one configuration prevails over the other
depends on the initial conditions.  Of course, integration is the more
likely outcome if the initial attitudes are highly cooperative, while
segregation will typically emerge from initial scenarios where guests
and hosts are highly hostile to each other.  This finding is also
consistent with sociological observations \cite{KOO10,PRI14} where the
attitude of the majority population is identified as a primary
determinant in minority segregation. 

We also find that
given the same social environment, a higher immigrant population ratio
results in segregation, while a larger total population will more
harmoniously absorb the same percentage of immigrants,
which agrees with previous Ising-type 
sociophysical models of immigrant integration \cite{GAL16b}.
Our results suggest accommodating newcomers 
in accordance with the host population.
Small, possibly rural, communities may not be optimal conduits to
integration compared to more populous cities, especially if the
percentage of migrants is large.  Examples of countries distributing
refugees in proportion to the population of receiving municipalities
include Denmark, from 1986 to 1998, Sweden, from 1987 to 1991, and the United Kingdom since 2000.
However, refugees were later found to
relocate to larger cities \cite{DAM09,STE11,WHI16}, attracted by the presence of
more co-ethnics, job opportunities and housing. Recent studies have
also observed higher segregation of immigrants in rural areas,
especially when the size of the migrant group is large and hosts are
hostile to guests \cite{KAN04,LIC16,LIC10}.

Our model does not include spatial dependence or
geographical factors in making and maintaining social connections.
For example, the turnover rate of social connections may be higher in
denser areas, leading to inhomogeneous timescales for attitude changes
in the network. We also do not consider the effects of
virtual connectivity, whereby internet connections may render spatial
dependence less relevant while also accelerating segregation as
finding co-cultural companions is facilitated in online venues.

Our model may be generalized by introducing a continuous influx of
immigrants, instead of assuming a fixed initial guest population.  A
continuous influx may allow us to include in-group interactions
between immigrants arriving at different times, and to study
cooperation or antagonism among them.  To further extend our model
across generations, since earlier immigrants and their descendants
eventually may be considered part of the native community, the $x_i^t
= 0$ barrier between hosts and guests must be relaxed in order to
allow for generational crossover between groups.  Similarly, long-term
attitude differences between hosts and guests can lead to open
conflict or violence that may curb socioeconomic rewards, including
in-group ones.  This mechanism would require higher order corrections
and feedback mechanisms that are currently not included in our work.

Moreover, our utility function $U_{i}^t$ carries the same form for
every node and penalizes those with too many connections. As a result,
all nodes converge towards an average number of connections, which is
not realistic, since actual social networks take on small-world
characteristics, with hub nodes having a large number of connections
\cite{WAT98}. Our model may be improved by introducing more nuanced
utility functions. For example, we may postulate that nodes with
larger socioeconomic utility are able to maintain a larger number of
connections, compared to those with lower utility, creating a
mechanism for hubs to emerge \cite{GAB13}.  All these
factors may influence the entire society and change host and guest
perceptions, in a positive or negative way.

\begin{acknowledgments}
  This work was made possible by support from grants ARO W1911NF-14-1-0472,
  ARO W1911NF-16-1-0165 (MRD), and NSF DMS-1516675 (TC).
\end{acknowledgments}

\bibliographystyle{apsrev}
\bibliography{integration}

\end{document}